# Segregation-driven cross-slip mechanism of shockley partials in the γ' phase of CoNi-based superalloys

Zhida Liang[1,*], Fengxian Liu[3], Xin Liu[2,*], Yang Li[4], Yinan Cui[2,*], Florian Pyczak[1]

1. Institute of Materials Research, Helmholtz-Zentrum hereon GmbH, 21502 Geesthacht, Germany
2. Applied Mechanics Lab., School of Aerospace Engineering, Tsinghua University, Beijing 100084, China
3. Department of Mechanics of Solids, Surfaces and Systems, Faculty of Engineering Technology, University of Twente, Drienerlolaan 5, 7522NB Enschede, Netherlands
4. School of Mechanics and Engineering Science, Shanghai University, Shanghai 200042, China

*Corresponding author: Zhida Liang, zhida.liang@outlook.com; Xin Liu, liuxin23@mails.tsinghua.edu.cn; Yinan Cui, cyn@mail.tsinghua.edu.cn

**Abstract**

In general, the cross-slip of superdislocations ($a/2\langle 011\rangle$) from {111} planes to {001} planes has been frequently observed in superalloys, which are accompanied by the formation of an antiphase boundary (APB) and driven by thermal activation. However, no prior studies have evidenced the occurrence of Shockley partial dislocation ($a/6\langle 112\rangle$) cross-slip within the γ′ phase of superalloys. In this work, we present a newly observed cross-slip phenomenon: the Shockley partial dislocations cross-slip from one {111} plane to another {111} conjugate plane, facilitated by the formation of a stair-rod dislocation in the ordered γ′ phase of CoNi-based superalloy. Compression tests were conducted at 850 °C with a strain rate of $10^{-4}$ $s^{-1}$. Defects such as stacking faults and dislocations, along with the associated chemical fluctuations, were characterized using high-resolution scanning transmission electron microscopy (HRSTEM) and energy-dispersive X-ray spectroscopy (EDS). Elemental segregation was found to reduce the activation energy required for cross-slip by decreasing the energies of stacking faults and dislocations. In addition to elemental segregation, local stress concentrations, arising from the combined effects of applied stress, shearing dislocations within the γ' phase, and dislocation pile-ups, also play a critical role in triggering cross-slip. The formation of sessile stair-rod dislocations via this newly identified Shockley partial cross-slip in the γ' phase is beneficial for enhancing the high-temperature deformation resistance of the alloy by increasing the critical resolved shear stress required for further plastic deformation.

# 1. Introduction

Superalloys are a class of high-performance materials engineered for use in extreme environments, particularly within aerospace and energy sectors (*Reed 2008, Sims et.al. 1987, Chatterjee et.al. 2021, Przybyla et.al. 2010*). Their exceptional mechanical strength at elevated temperatures is primarily attributed to the presence of ordered L1$_2$-structured γ′ phases dispersed within a disordered face-centered cubic (fcc) γ matrix (*Li et.al. 2021, Feng et.al. 2010*). The γ′ phase, typically of Ni$_3$(Al,Ti) composition, plays a vital role in strengthening the alloy by impeding dislocation motion. In particular, under sustained mechanical loading at high temperatures, the interactions between dislocations and γ′ phases become critical in governing the superalloy's plastic deformation mechanisms (*Wu et.al. 2017, Nganbe et.al. 2009, Hou et.al. 2024*).

During the deformation process, the γ′ phases prevent the movement of $a/2\langle 011 \rangle$ matrix dislocations (where *a* represents lattice parameter), a mechanism known as precipitation strengthening (*Chang et.al. 2018, Wang et.al. 2025, Zhang et.al. 2024*). When dislocations shear the γ′ phases, various planar faults may form, including anti-phase boundaries (APBs), superlattice intrinsic stacking faults (SISFs), superlattice extrinsic stacking faults (SESFs), complex stacking faults (CSFs), and deformation twins (*Borovikov et.al. 2023, Borovikov et.al. 2025, Lenz et.al. 2019, Titus et.al. 2015, Barba et.al. 2017, Kovarik et.al. 2009, Smith 2016, Tsuno at.al. 2008*). These stacking faults significantly influence the deformation behaviour of superalloys, contributing to either strengthening or softening, depending on their configuration and local chemistry with elements segregation playing a critical role. While stacking faults often appear individually, they can also interact to form complex configurations such as V-, T-, and X-like structures during high-temperature deformation (*Qi at.al. 2016, Lu at.al. 2020*), see **Fig. 1**. These interactions can enhance creep resistance. For instance, V-like stacking fault configurations strengthen the γ′ phase by forming sessile stair-rod dislocations (*Qi at.al. 2016*), and T-like configurations create coherent stacking fault interfaces that block further dislocation motion (*Lu at.al. 2020*).

At elevated temperatures, in addition to dislocation glide and shearing, dislocation climb and cross-slip are also activated and become key contributors to plastic deformation. Cross-slip, a process in which a screw dislocation transfers from its original slip plane to a secondary one, is a fundamental mechanism in fcc metals such as aluminum (*Jin at.al. 2011, Hazif at.al. 1975, Watanabe at.al. 2021*), copper (*Bonneville at.al. 1988, Rasmussen at.al. 1997, Rao at.al. 2010*), and nickel (*Kang et.al. 2014, Kuykendall et.al. 2020, Wen at.al. 2004*). In the context of superalloys, cross-slip modes of $a/2\langle 011 \rangle$ full dislocations have been well documented. First, the cross-slip of $a/2\langle 011 \rangle$ superdislocations from the {111} to the {001} plane has been associated with the formation of Kear-Wilsdorf (KW) locks (*Lin et.al. 1990, Milligan et.al. 1989, Rai et.al. 2021*), owing to the lower APB energy in the {001} plane. This process is particularly important in superalloys, as it contributes directly to the anomalous increase in strength at high temperatures. Second, $a/2\langle 011\rangle$ full dislocation cross-slip is frequently observed in the γ matrix during creep deformation (*Zhang et.al. 2005*), facilitating the development of dislocation networks during creep rafting. Third, $a/2\langle 011\rangle$ dislocation cross-slip at γ/γ′ interfaces has been experimentally linked to SISF formation via the Kear mechanism (*León-Cázares et.al. 2022*).

Despite these insights, a critical gap remains that all experimentally confirmed cross-slip events in superalloys have involved full dislocations with Burgers vectors of $a/2\langle 011\rangle$. To date, no direct experimental evidence has been reported for the cross-slip of Shockley partial dislocations ($a/6\langle 112\rangle$) within ordered γ′ phases. This is particularly important given the longstanding debate over the formation mechanisms of V-like stacking fault configurations, such as SISF-SISF, SESF-SESF, and SISF-SESF. While these configurations are traditionally attributed to the interaction of two leading partial dislocations forming Lomer-Cottrell (L-C) locks (*Qi et.al. 2016, Behjati et.al. 2011, Zeng et.al. 2024, Zhou et.al. 2017*), the likelihood of such events is very low. An alternative and increasingly considered mechanism involves the cross-slip of Shockley partial dislocations. However, both direct experimental evidence and clear theoretical analysis of their energetics have been lacking, hindering development of predictive models for γ′ phases deformation mechanism at high temperatures.

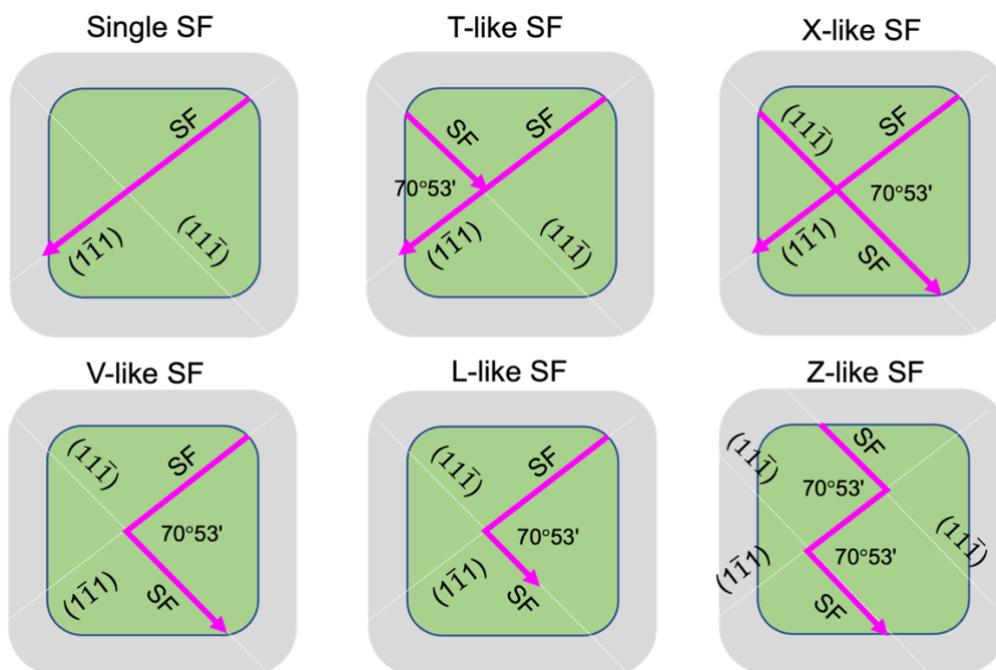

**Fig.1** Schematic diagram about various stacking fault (SF) interaction morphologies, including single, V-, T-, L-, Z-, and X-like SF features.

In this study, we report the first direct experimental observation of Shockley partial dislocation cross-slip occurring within the interior of ordered γ′ phases in a CoNi-based superalloy. Detailed analysis of novel SF configurations, including L- and Z-shaped structures, reveals that cross-slip is closely linked to the formation of sessile stair-rod dislocations. Under dislocation cross-slip, V-shaped SF interactions appear to represent a transient state between the L- and Z-shaped configurations, see **Fig. 1**. These observations provide compelling evidence for a unique cross-slip mechanism of Shockley partials in ordered γ′ phases. To further substantiate this mechanism, a simplified analytical model is proposed to describe the thermodynamic and kinetic conditions facilitating cross-slip, emphasizing the critical roles of elemental segregation and local stress concentration. By comparing this mechanism with those in pure fcc metals and solid-solution-strengthened alloys, our findings offer new insights into the distinct cross-slip behaviour of ordered systems. Moreover, understanding these stacking fault interactions and the associated stair-rod dislocations offers a promising strategy for improving the high-temperature deformation resistance of advanced superalloys.

## 2. Experiments

The investigated alloy is a γ'-hardened polycrystalline CoNi-based superalloy, designated as CoNi-2Mo1W. Its nominal chemical composition is $Co_{35}Ni_{15}Cr_5Al_5Ti_2Mo_1W_{0.1}B$ (at.%), while the actual composition, determined via in Energy-dispersive X-ray spectroscopy (EDS) analysis in scanning electron microscopy (SEM-EDS), is presented in **Table 1**. The alloy was synthesized by vacuum arc melting under an argon atmosphere, forming a 70 g ingot on a water-cooled copper hearth. To ensure chemical homogeneity, the material underwent at least seven remelting cycles. Following casting, the alloy was homogenized at 1250 °C for 24 hours, aged at 900 °C in air for 220 hours, and subsequently air-cooled to room temperature.

**Table 1.** The measured composition (at.%) of the superalloy CoNi-2Mo1W by SEM-EDS.

| Elements | Co | Ni | Cr | Al | Ti | Mo | W |
|---|---|---|---|---|---|---|---|
| Fraction | 37.4 | 34.3 | 15.6 | 4.6 | 5.3 | 1.9 | 0.9 |

Compression tests were conducted to determine the 0.2% proof stress of the alloy using a strain-controlled, closed-loop testing machine (MTS 810, MTS Systems Corporation) at 850 °C with a strain rate of $10^{-4}$ s$^{-1}$. Cylindrical compression specimens, each with a gauge length of 7.5 mm and a diameter of 5 mm, were machined from the standard heat-treated material. Deformation was monitored using an extensometer attached over a 21 mm gauge length, with the measured extension serving as the feedback parameter.

For transmission electron microscopy (TEM) sample preparation, 3 mm discs were extracted from the heat-treated specimens by sectioning at a 45° angle along the axial direction of the cylindrical sample. These discs were mechanically polished to a thickness of approximately 75 µm before being further thinned to electron transparency using a twin-jet electro-polishing unit. The final thinning process was carried out in Struers A3 electrolyte at -38 °C with an applied voltage of 32 V. Low magnification annular bright-field scanning transmission electron microscopy (ABF-STEM) analysis was performed using a Thermo Fisher Scientific Talos F200i operated at 200 kV.

STEM-EDS analysis was performed using a double-corrected Thermo Fisher Themis-Z TEM, equipped with a four-quadrant Super-X EDS detector. High-resolution scanning transmission electron microscopy (HRSTEM) images and EDS maps were acquired at an accelerating voltage of 300 kV to analyze elemental distribution around stacking fault defects. The high-angle annular dark-field (HAADF) detector had a collection angle range of 51-200 mrad, with a camera length of 145 mm. The acquisition parameters for EDS mapping and HAADF-HRSTEM imaging were as follows: the convergence angle was 30 mrad (C2 = 70 µm), and the probe current was approximately 300 pA for both HAADF-HRSTEM images and EDS maps. In addition, the High resolution HAADF-STEM images were analysed by center of symmetry (COS) method with MATLAB codes (*Messé et.al. 2014*).

3. Results

The shearing of Shockley partial dislocations in the γ′ phase is accompanied by the formation of SFs. Accordingly, our investigation primarily focuses on the configurations of SFs and the elemental distribution in their vicinity.

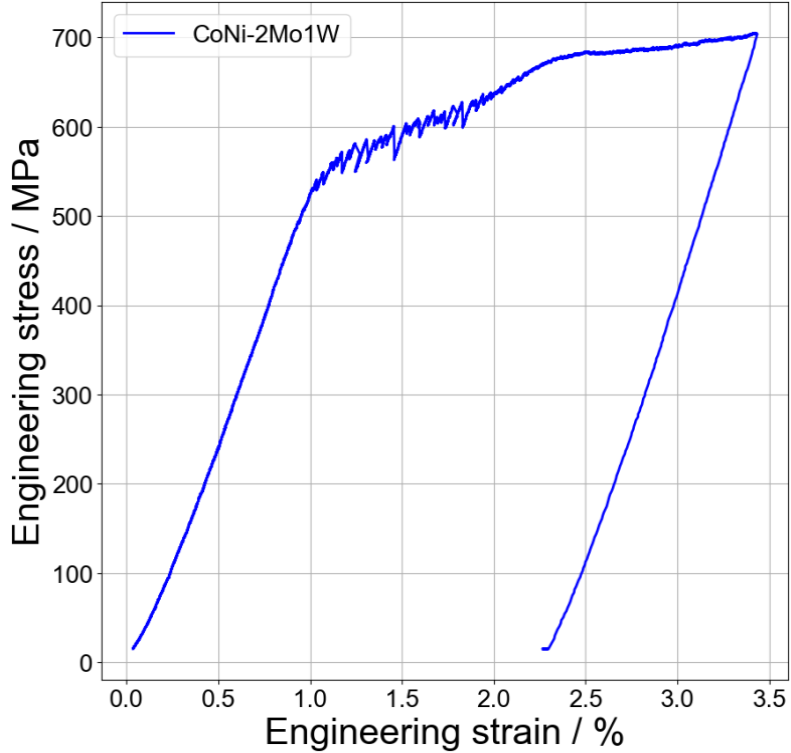

**Fig. 2** Engineering stress-strain curve of superalloy CoNi-2Mo1W at 850 °C.

**Fig. 2** presents the engineering stress-strain curve of the CoNi-2Mo1W superalloy at 850 °C. The compression test was interrupted at a strain of 3.4% under a stress of 700 MPa. The 0.2% yield stress was measured to be 572 MPa. **Figs. 3(a)** and **(b)** show ABF-STEM images of the deformation substructures (at ~3.4% compression), corresponding to **Fig. 2**. The micrographs (beam direction close to the $\boldsymbol{\mu} = [011]$ direction and $\vec{g} = (200)$) shows the $(11\bar{1})$ and $(1\bar{1}1)$ planes edge-on, and the SFs lying in these two planes show up as thin grey lines inside the γ′ phase.

Various SF interactions, including single, V-, T-, L-, Z-, and X-like features (named based on their two-dimensional appearances), are observed and some are marked with white arrows in **Figs. 3(a)** and **(b)**, which morphologies were illustrated in **Fig. 1**. These interactions typically occur at angles of ~ 70.53° between intersecting SFs. A total of 334 SFs were counted within an area of ~ 46 μm² in a [011] oriented foil and categorized into six types based on the nature of the intersections. The fraction of isolated single SFs occupied 69.5% around, see **Table 2** and **Fig. 3(c)**. Among the observed SF interactions, T-like and X-like types were the most frequent, with fractions of 13.5% and 9.2%, respectively, followed by V-like (5.1%), L-like (1.8%), and Z-like (0.9%) types. The V-, T-, and X-like SF interactions have been previously discussed in detail by Lu et al. (*Lu et al. 2020*), where it was suggested that such interactions enhance creep resistance compared to isolated SFs. However, detailed reports on L-like and Z-like SF interactions have been lacking. In the present study, we focus on two examples illustrating L-like and Z-like SF interactions, characterized by HRSTEM imaging and EDS mapping. Furthermore, the identification of L-like and Z-like interactions provides direct evidence for the cross-slip of Shockley partial dislocations within the γ' phase. This finding further supports the previous assertion that the V-like SF interaction, higher probability than L-like and Z-like, could also arise from cross-slip.

**Table 2** Number and fraction of different SF types in the γ′ phase in ~ 46 um$^2$ area of target grain with 3.4% strain.

| SF type | Single | Interaction | | | | | Total |
|---|---|---|---|---|---|---|---|
| | | T-like | X-like | V-like | L-like | Z-like | |
| Number | 232 | 45 | 31 | 17 | 6 | 3 | 334 |
| Fraction | 69.5 % | 13.5 % | 9.2 % | 5.1 % | 1.8 % | 0.9 % | 100 % |

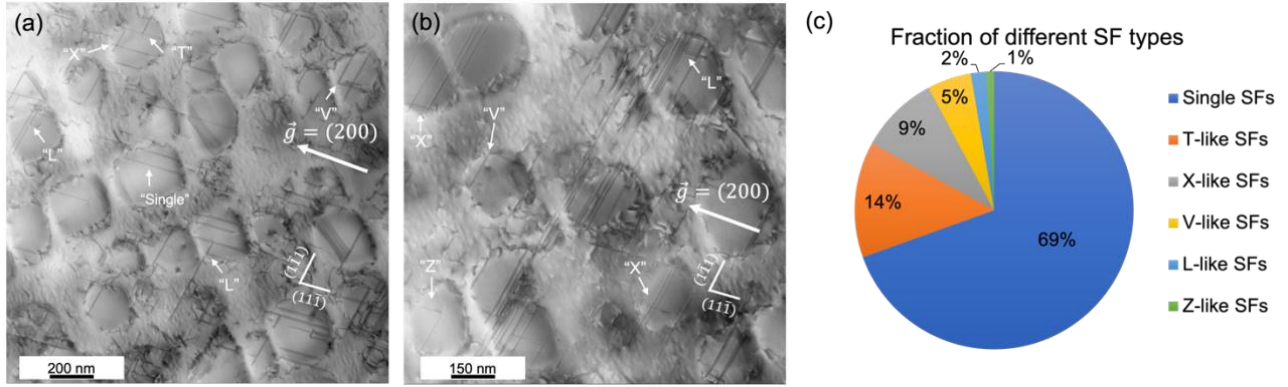

**Fig. 3** (a) and (b) ABF-STEM images of deformation substructures (compressed by 3.4%.) of alloy CoNi-2Mo1W corresponding to **Fig. 2**. Electron beam is parallel to [011] crystal direction. (c) Pie chart illustrating the fraction of various SF types.

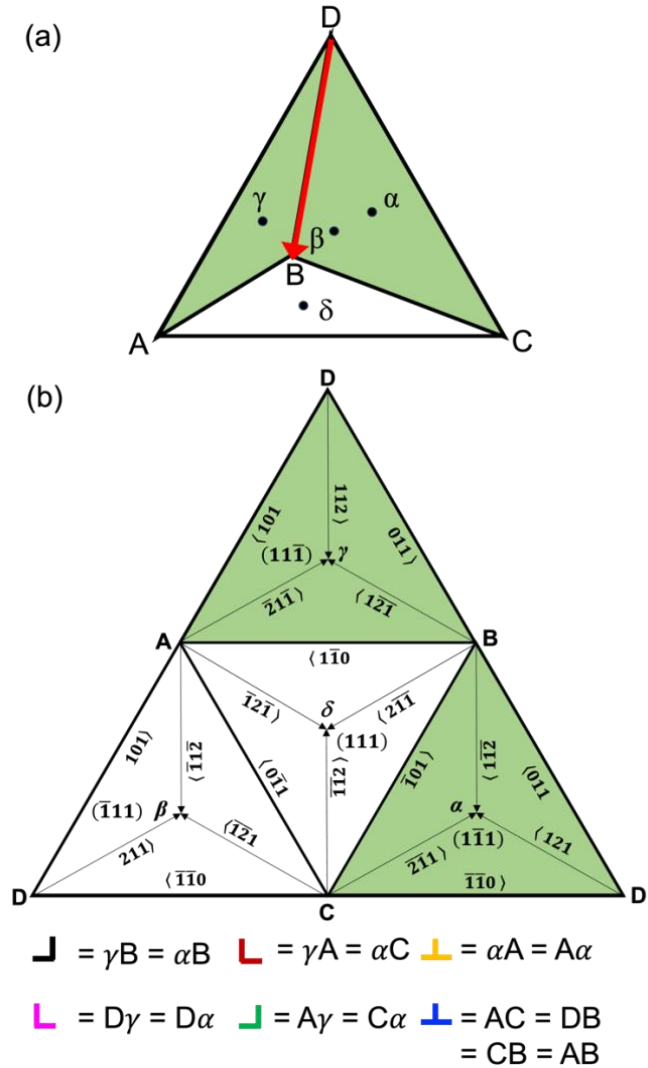

**Fig. 4** (a) Thompson tetrahedron. **DB** is electron beam direction in TEM. The planes **ABD** and **CBD** are $(11\bar{1})$ and $(1\bar{1}1)$, respectively. (b) Coordinate system of unfolded Thompson tetrahedron, where the displayed vectors only indicate the directions. **DB** is equal to $a/2[011]$ and **Dγ** is equal to $a/6[112]$. All vectors mentioned in this work refer to the coordinate system in (b). (The black and pink signs represent 30° Shockley partial dislocation, the red and green signs represent 90° Shockley partial dislocation, the orange sign represents 90° Frank partial dislocation and the blue sign represents $a/2\langle110\rangle$ full dislocation)

For clarity, all descriptions of stacking faults and dislocations in this study are labelled with reference the coordinate system superimposed on a Thompson tetrahedron shown in **Fig. 4**. **ABC**, **ABD**, **ACD** and **BCD** are four conjugate {111} planes on which stacking faults form. Vectors that are denoted by a pair of Roman letters, Roman-Greek letters or Greek letters represent Burgers vectors of full dislocations, super partial dislocations, Frank, Shockley and stair-rod partial dislocations, respectively. For instance, **DB** represents the Burgers vector of $a/2[011]$ full dislocation, α**A** represents the Burgers vector of $a/3\,[1\bar{1}1]$ 90° Frank partial dislocation, **Dγ** corresponds to the Burgers vector of $a/6[112]$ 30° Shockley partial dislocation, **Aγ** represents the Burgers vector of $a/6[\bar{2}1\bar{1}]$ 90° Shockley partial dislocation and **αγ** denotes the Burgers vector of $a/6\,[0\bar{1}1]$ 90° stair-rod dislocation ($a$ = 0.356 nm in this work).

### 3.1 L-like SF interaction

**Fig. 5(a)** shows a single γ' precipitate containing several individual SFs and a L-like SF interaction configuration, schematically represented in red in **Fig. 5(b)**. To investigate the L-like SF configurations, the small boxes labelled i, ii, and iii in **Fig. 5(a)** are further examined by employing HRSTEM. The results, shown in **Figs. 5(c), (d) and (e)** respectively, reveal that the L-like SF interaction configuration consists of two SISFs and one SESF. Detailed investigation indicates that the leading edge of the configuration comprises two Shockley partial dislocations (**Fig. 5(c)**), while the trailing edge contains one Shockley partial (**Fig. 5(e)**). All Shockley partial dislocations are aligned along the electron beam direction, **μ** = [011], corresponding to the **DB** direction in the tetrahedron illustrated in **Fig. 4**.

Based on the Burgers vector analysis criteria, the Burgers vector is defined as the vector from the finish point (F) to the start point (S) of the Burgers circuit. Clockwise Burgers circuits were drawn around the leading and trailing edges, as well as around the SESF-SISF intersection, as shown in **Figs. 5(f), 5(g), and 5(h)**. For the leading partials, where the SISF is bounded between two leading partials in crystal plane $(11\bar{1})$ (**Fig. 5(f)**), the projected Burgers vectors are determined to be $b_{1,p} = F_1S_1 = a/12[\bar{2}1\bar{1}]$ and $b_{2,p} = F_2S_2 = a/6[\bar{2}1\bar{1}]$ in the projection plane. The actual Burgers vectors, deduced from the projection, are $b_1 = a/6[\bar{1}21]$ and $b_2 = a/3[\bar{1}21]$, corresponding to two **γB** dislocations in the Thompson tetrahedron. Given that the angle between the dislocation line **DB** and the Burgers vector **γB** is approximately 30°, the two leading partials are identified as 30° mixed-type Shockley partials. At the intersection point, three Burgers circuits were analyzed, yielding projected Burgers vectors $b_{3,p} = F_3S_3 = a/12[\bar{2}1\bar{1}]$, $b_{4,p} = F_4S_4 = a/12[2\bar{1}1]$ and $b_{5,p} = F_5S_5 = a/6[\bar{2}1\bar{1}]$. Accordingly, at the SESF-SISF intersection shown in **Fig. 5(d)**, one dislocation with $b_{3,4} = b_{3,p} + b_{4,p} = a/6[0\bar{1}1]$, is identified as a 90° stair-rod partial (**αγ**), while the other dislocation with $b_5 = a/6[\bar{2}1\bar{1}]$ corresponds to a 90° Shockley edge partial (**Aγ**).

Further explanations for distinguishing between 30° mixed-type and 90° edge-type Shockley partials, as well as the method for determining the real Burgers vectors from the projected ones in HRSTEM images, are provided in **Supplementary Materials 1 Fig. S1**. As shown in **Fig. 5(h)**, the trailing partial, where a SISF is bounded between the trailing partial and the intersection point in crystal plane $(1\bar{1}1)$, has a projected Burgers vector of $b_{6,p} = F_6S_6 = a/12[21\bar{1}]$ in the projection plane. From this, the real Burgers vector is deduced to be $b_6 = a/6[121]$ corresponding to the $D\alpha$ dislocation in the Thompson tetrahedron, identified as a 30° Shockley partial. The detailed discussion for L-like SF interaction, including dislocation cross-slip mechanism, is shown in **section 4.1**.

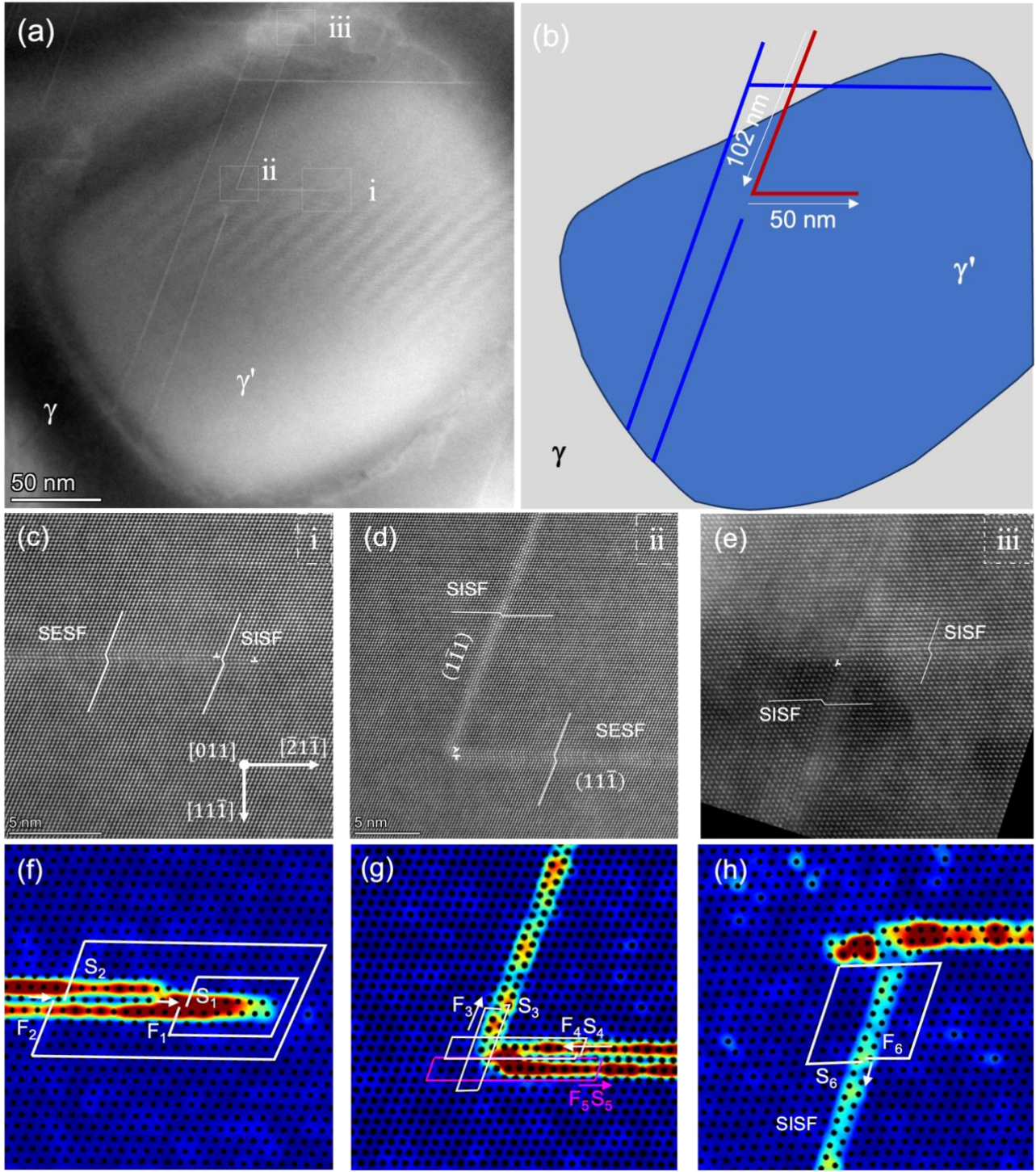

**Fig. 5** (a) HAADF-STEM image of single γ′ particle with L-like SF interaction. (b) The schematic illustration of the configuration in **Fig. 5(a)**. (c) Atomic resolution HAADF-STEM image with defects of leading partial dislocations (LPDs) and SESF. (d) Atomic resolution HAADF-STEM image with the SESF-SISF intersection. (e) Atomic resolution HAADF-STEM image with one trailing dislocation and SISF. (f), (g) and (h) COS images for **Figs. 5(c), (d)** and **(e)**, respectively.

## 3.2 Z-like SF interaction

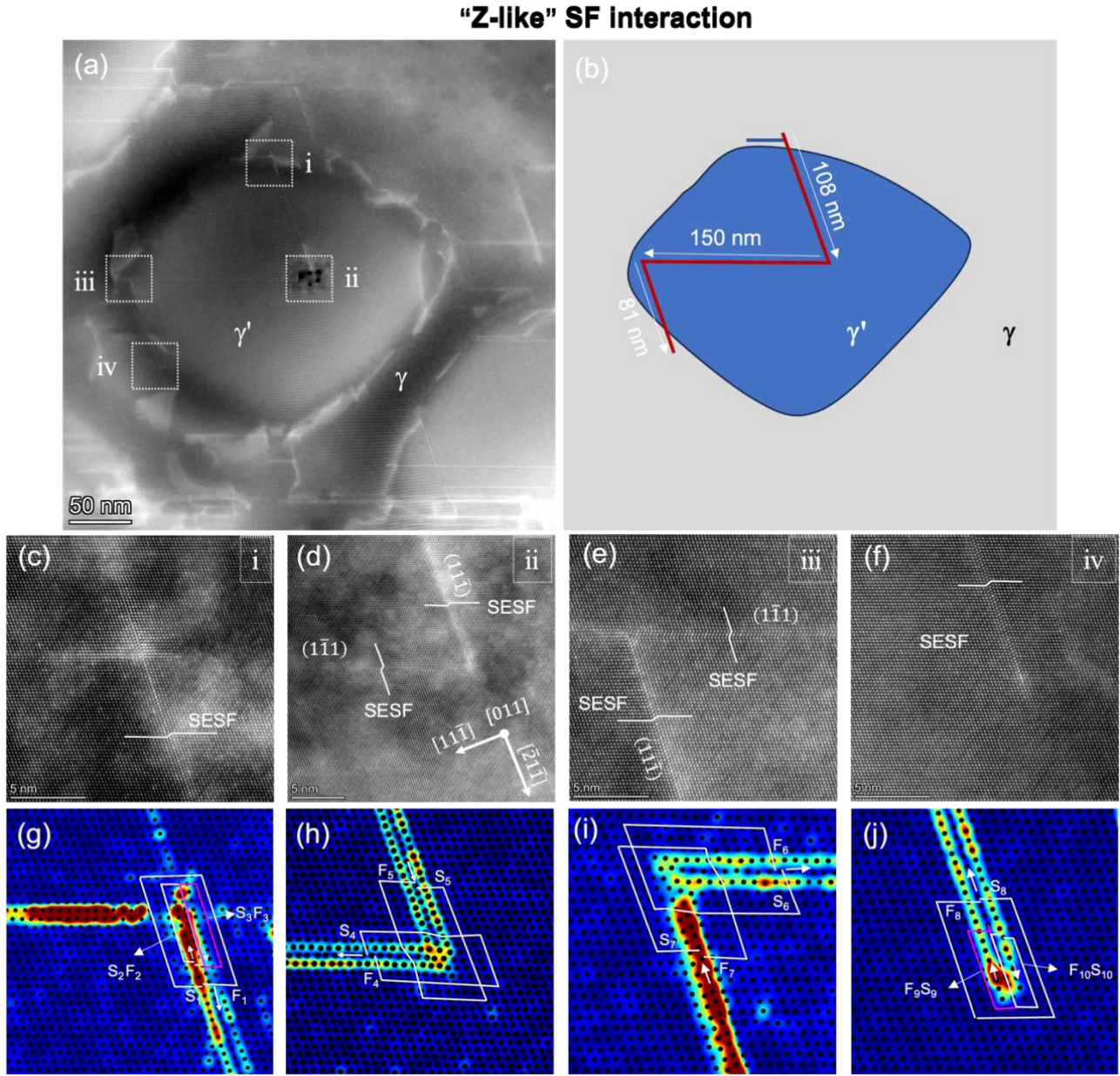

**Fig. 6** (a) HAADF-STEM image of single γ′ particle with Z-like SF interations. (b) The relevant diagram for **Fig. 6(a)**. (c) Atomic resolution HAADF-STEM image with defects of trailing dislocation and SESF. (d) Atomic resolution HAADF-STEM image with the first SESF-SESF intersection (The first intersection point). (e) Atomic resolution HAADF-STEM image with the second SESF-SESF intersection (The second intersection point). (f) The atomic resolution HAADF-STEM image with defects of LPDs and SESF. (g), (h), (i) and (j) COS images for **Fig. 6(c), (d), (e)** and **(f)**, respectively.

**Fig. 6(a)** shows a single γ' precipitate containing a Z-like SF interaction configuration, which is highlighted with red in **Fig. 6(b)**. To investigate the Z-like SF configurations, the small boxes labelled i, ii, iii and iv in **Fig. 6(a)** are further examined by employing HRSTEM. HRSTEM analysis reveals that the Z-like configuration consists of three SESFs, as shown in **Figs. 6(c), (d), (e),** and **(f)**. It is observed that both the leading and trailing edges are each bounded by two partial dislocations, with all dislocation lines aligned along the electron beam direction, $\boldsymbol{\mu}$ = [011].

Clockwise Burgers circuits were drawn around the two edges and the two SESF-SESF intersections, as shown in **Figs. 6(g), (h), (i),** and **(j)**. The trailing partial in crystal plane $(11\bar{1})$, shown in **Fig. 6(g)**, exhibits three Burgers circuits with three projected Burgers vector of $\mathbf{b}_{1,p} = \mathbf{F}_1\mathbf{S}_1 = a/12[\bar{2}1\bar{1}]$, $\mathbf{b}_{2,p} = \mathbf{F}_2\mathbf{S}_2 = a/12[2\bar{1}1]$ and $\mathbf{b}_{3,p} = \mathbf{F}_3\mathbf{S}_3 = a/6[\bar{2}1\bar{1}]$ in the projection plane. The corresponding real Burgers vector are deduced to be $\mathbf{b}_1 = a/6[\bar{1}21]$ (γ**B**), $\mathbf{b}_2 = a/6[112]$ (**D**γ) and $\mathbf{b}_3 = a/6[\bar{2}1\bar{1}]$ (**A**γ). Given the 30° angle between the dislocation line and Burgers vector, the trailing partial **D**γ is identified as a 30° mixed Shockley partial and **A**γ is idensitfied as a 90° edge Shockley partial. Actually, $\mathbf{b}_1 = \mathbf{b}_2 + \mathbf{b}_3$ means the trailing partials consists of a 90° edge Shockley partial (**A**γ) and a 30° mixed Shockley partial (**D**γ).

At the first intersection point, shown in **Fig. 6(h)**, two projected Burgers circuits were analyzed, giving $\mathbf{b}_{4,p} = \mathbf{F}_4\mathbf{S}_4 = a/6[21\bar{1}]$ and $\mathbf{b}_{5,p} = \mathbf{F}_5\mathbf{S}_5 = a/6[\bar{2}1\bar{1}]$. The sum of these vectors yields $\mathbf{b}_{4,5} = \mathbf{b}_{4,p} + \mathbf{b}_{5,p} = a/3[01\bar{1}]$, corresponding to two 90° stair-rod partial dislocations, denoted as **2γα**. Similarly, at the second intersection point, shown in **Fig. 6(i)**, two projected Burgers circuits were determined: $\mathbf{b}_{6,p} = \mathbf{F}_6\mathbf{S}_6 = a/6[\bar{2}11]$ and $\mathbf{b}_{7,p} = \mathbf{F}_7\mathbf{S}_7 = a/6[2\bar{1}1]$. Their sum gives $\mathbf{b}_{6,7} = \mathbf{b}_{6,p} + \mathbf{b}_{7,p} = a/3[0\bar{1}1]$, corresponding to two 90° stair-rod partial dislocations, denoted as **2αγ**.

Finally, the leading partial dislocations in crystal plane $(11\bar{1})$, shown in **Fig. 6(j)**, contains three Burgers circuits with corresponding projected Burgers vectors: $\mathbf{b}_{8,p} = \mathbf{F}_8\mathbf{S}_8 = a/12[2\bar{1}1]$, $\mathbf{b}_{9,p} = \mathbf{F}_9\mathbf{S}_9 = a/6[2\bar{1}1]$ and $\mathbf{b}_{10,p} = \mathbf{F}_{10}\mathbf{S}_{10} = a/12[\bar{2}1\bar{1}]$, all within the projection plane. The true Burgers vectors are determined as: $\mathbf{b}_8 = a/6[112]$ (denoted **D**γ), $\mathbf{b}_9 = a/6[2\bar{1}1]$ (γ**A**) and $\mathbf{b}_{10} = a/6[\bar{1}21]$ (γ**B**). Here, γ**B** is identified as a 30° mixed Shockley partial, and γ**A** as a 90° edge Shockley partial. Notably, the vector relation $\mathbf{b}_8 = \mathbf{b}_9 + \mathbf{b}_{10}$ confirms that the leading partial consists of a 90° edge Shockley partial (γ**A**) and a 30° mixed Shockley partial (γ**B**).

The analysis for leading and trailing partials in the Z-like SF interaction is consistent with the SESF mode (SESF-1) in **Appendix 1**. The detailed discussion for Z-like SF interaction, including dislocation cross-slip and *cross-split* mechanism, is shown in **section 4.1**. **Fig. 7** summarised the detailed configurations of L-like and Z-like SF interactions, including the trailing and leading partial dislocations, based on analysis in **sections 3.1** and **3.2**.

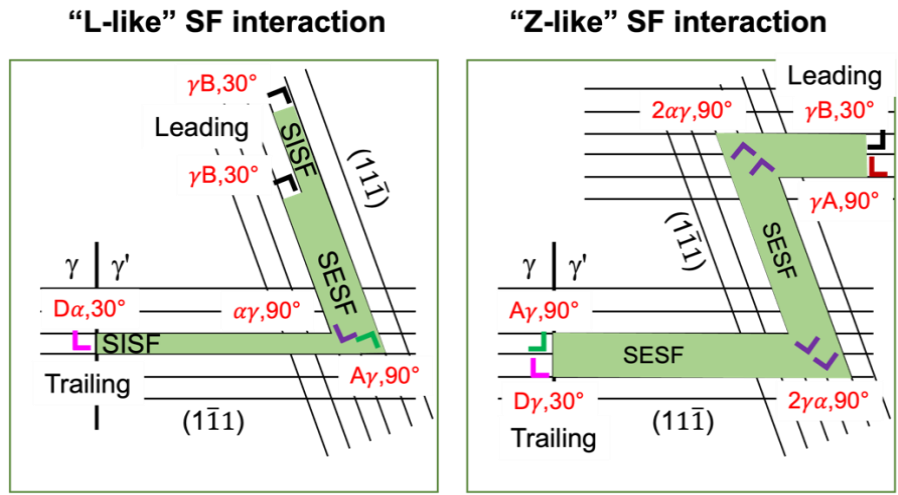

**Fig. 7** Schematic model of L-like and Z-like SF interactions based on analysis in **sections 3.1** and **3.2**.

**3.3. High resolution EDS analysis and segregation behavior of SF interactions**

To investigate the chemical fluctuations around defects, including stacking faults and dislocations, high-resolution STEM-EDS analysis was performed. The corresponding results are presented below.

**Table 3.** Local compositions (at.%) and segregation magnitude of the surrounding γ′ phase, LPD, Stair-rod dislocation, SISF and SESF in L-like SF interaction and Z-like SF interaction, respectively.

| Element | γ′ phase | L-like SF interaction | | | | Z-like SF interaction | |
|---|---|---|---|---|---|---|---|
| | | SISF | SESF | LPD | Stair-rod | SESF | Stair-rod |
| **Co** | 24.9±0.2 | 30.2±0.4 | 28.2±0.0 | 28.0±0.3 | 33.2±0.1 | 31.6±0.2 | 34.8±0.0 |
| **Ni** | 50.8±0.3 | 43.5±0.4 | 45.0±0.2 | 45.8±0.2 | 41.4±0.0 | 41.5±0.1 | 38.1±0.5 |
| **Cr** | 2.9±0.0 | 5.3±0.0 | 5.4±0.1 | 6.5±0.2 | 7.7±0.0 | 5.7±0.0 | 8.5±0.1 |
| **Al** | 7.4±0.2 | 6.0±0.1 | 6.5±0.2 | 6.9±0.6 | 5.3±0.1 | 6.4±0.1 | 5.8±0.2 |
| **Ti** | 12.0±0.2 | 11.3±0.1 | 11.4±0.2 | 10.3±0.4 | 9.2±0.1 | 10.6±0.0 | 9.7±0.0 |
| **Mo** | 0.8±0.0 | 2.1±0.1 | 1.7±0.0 | 1.4±0.2 | 1.7±0.1 | 2.4±0.1 | 1.6±0.2 |
| **W** | 1.3±0.0 | 1.7±0.0 | 1.7±0.1 | 1.2±0.0 | 1.5±0.1 | 1.8±0.1 | 1.5±0.1 |

**Figs. 8 and 9** present high-resolution HAADF-STEM images and the corresponding EDS analyses (at.%) around the SISF-SESF intersection of the L-like SF interaction and the SESF-SESF intersection of the Z-like SF interaction, respectively. In both cases, similar segregation behavior is observed: Co, Cr, Mo, and W are enriched at the SFs, while Ni and Al are depleted, with Ti remaining nearly uniformly distributed. At the SF intersections (i.e., cross-slip points), a comparable segregation trend is observed, although the Z-contrast is higher, as shown in **Fig. 8**. Notably, the intersection of two SESFs exhibits even stronger Co and Cr enrichment and more pronounced depletion of Ni, Ti, and Al compared to individual SFs. Segregation at the leading partial dislocation (LPD) is even more significant, particularly for Co and Cr, as summarized in **Table 3**. Similarly, the HAADF-STEM image and EDS analysis around the first cross-slip point in **Fig. 9** show elemental distributions consistent with those observed in **Fig. 8**. The strong depletion of γ′-forming elements (Ni, Al, and Ti) around the cross-slip point suggests that the composition near the edge dislocation at these intersections tends toward local chemical disorder.

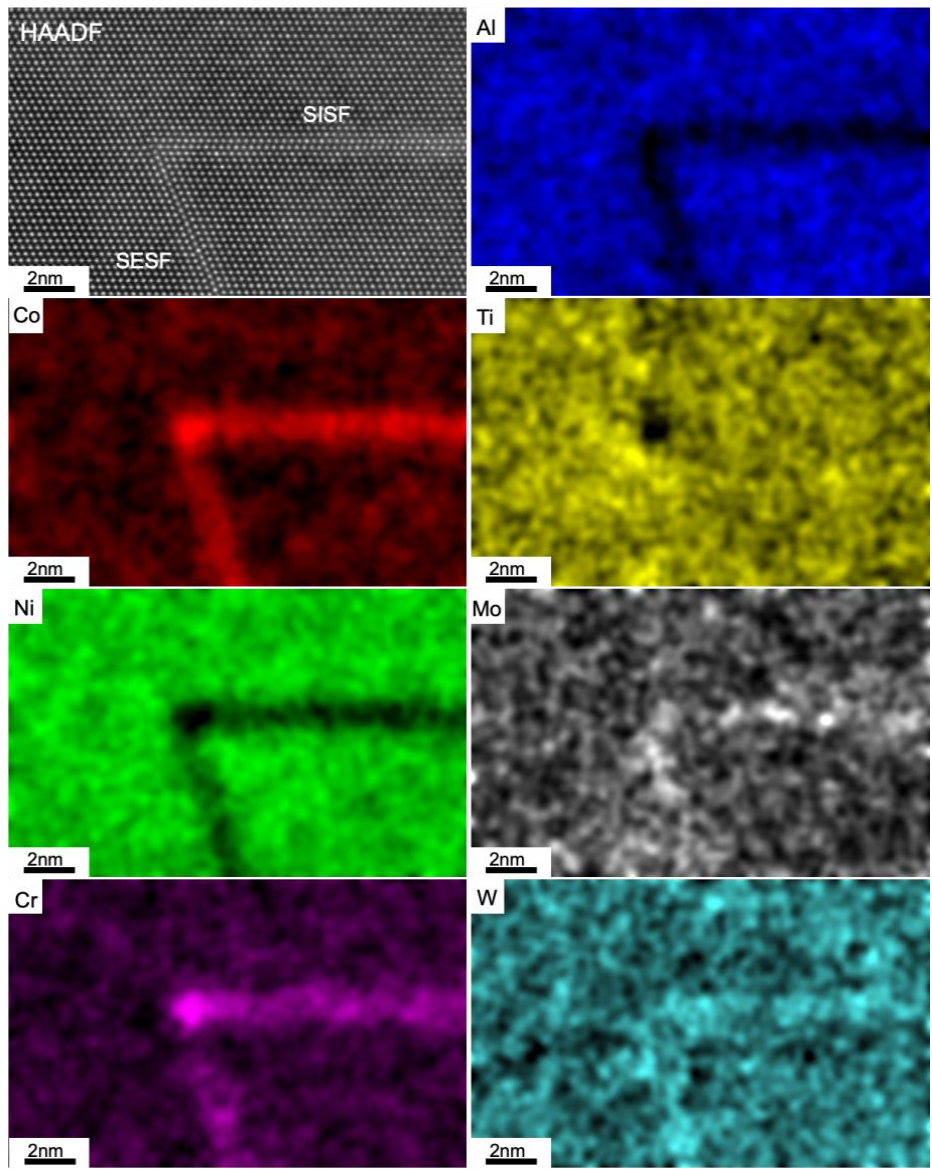

**Fig. 8** High resolution HAADF-STEM image and EDS elemental mapping (at.%) of the intersections with SESF, SISF and stair-rod edge dislocation, from the same position as shown in **Fig. 5(d)**.

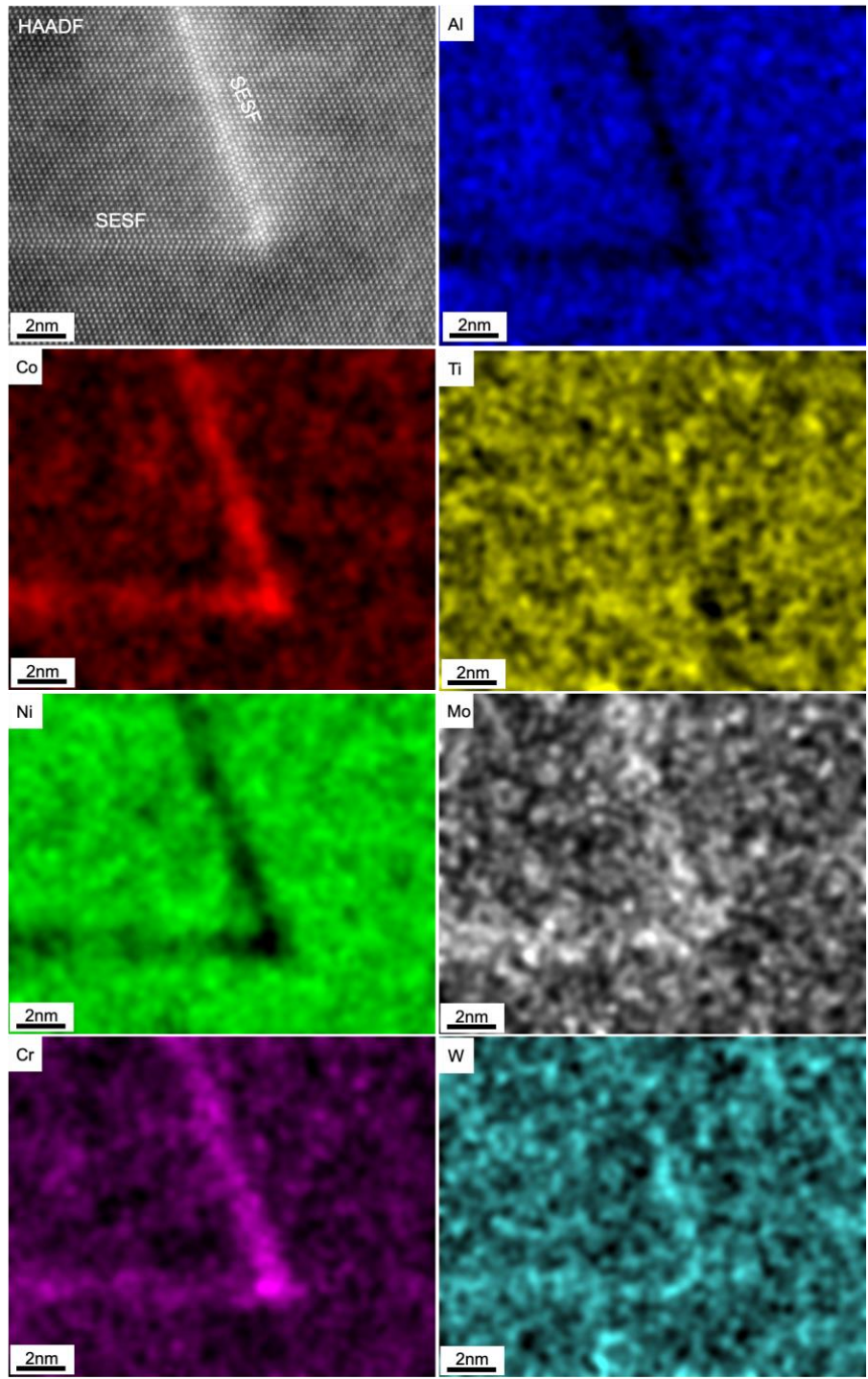

**Fig. 9** High resolution HAADF-STEM image and EDS elemental mapping (at.%) of the intersection with SESF, SISF and stair-rod edge dislocation, from the same position as shown in **Fig. 6(d)**.

## 4. Discussion

In this work, we focus on two characteristic SF interaction configurations: L-like and Z-like SF interactions, including their associated SFs and dislocation structures. We identify that the cross-slip or *cross-split* of Shockley partial dislocations predominantly occurs at the intersection points of these two types of interactions. Based on this observation, we carried out a detailed investigation as follows.

First, we analyze the cross-slip mechanism of 30° Shockley partial dislocations within the L-like SF interaction. Second, we propose two distinct interaction modes to explain the Z-like SF configuration. Third, we present a simplified mathematical model to describe how elemental segregation alters defect energies, specifically those of SFs and dislocations, and how this modification directly influences the activation energy required for cross-slip or *cross-split* of Shockley partial dislocations. Finally, we examine the role of local stress fields surrounding the γ′ phase as an additional critical factor in determining the threshold stress necessary to activate cross-slip or *cross-split* processes.

**4.1. Stacking fault type and dislocation dissociation/reaction/cross-slip/*cross-split***

In order to understand the SESF configurations in superalloys, **Appendix 1** summarizes three distinct SESF shearing modes, SESF-1, SESF-2, and SESF-3, observed within the γ′ phase of superalloys. All three types of SESF configurations are enclosed by four Shockley partial dislocations. SESF-3 has previously been deduced to originate from synthetic reactions between two CISFs (*Karpstein et.al. 2023, Vorontsov et.al. 2012, Smith et.al. 2015, Borovikov et.al. 2024*). In the present work, we identify two additional modes, SESF-1 and SESF-2, which can also be rationalized as products of synthetic reactions between two CISFs. By element segregation and reordering, two CISFs eventually evolved into one SESF. Based on the Thompson tetrahedron, the configurations of the three reported SISFs are illustrated in **Figs. 10(a), (b), and (c)**. SISF-1 is enclosed by two 30° mixed-type Shockley partials (*Vorontsov et.al. 2012*), whereas SISF-2 and SISF-3 are enclosed by a combination of a 30° mixed-type and a 90° edge-type Shockley partials (*Smith et.al. 2018*).

**Figs 10. (a), (b), and (c)** illustrate the dissociation of different types of perfect dislocations into SISFs enclosed by two partials. **Figs. 10(d), (e), and (f)** schematically show the formation pathways of SESF-1, SESF-2, and SESF-3 through synthetic reactions between those different types of CISFs, having similar substructures with SISFs, as illustrated in **Figs. 10(a), (b), and (c)**. This mechanism provides a consistent explanation for how an SESF can originate from the reaction between a pure screw perfect dislocation **DB** and a 60° mixed perfect dislocation **AB** or between two 60° mixed perfect dislocation (**AB** and **DA**) on the $(11\bar{1})$ plane, denoted as **DAB**.

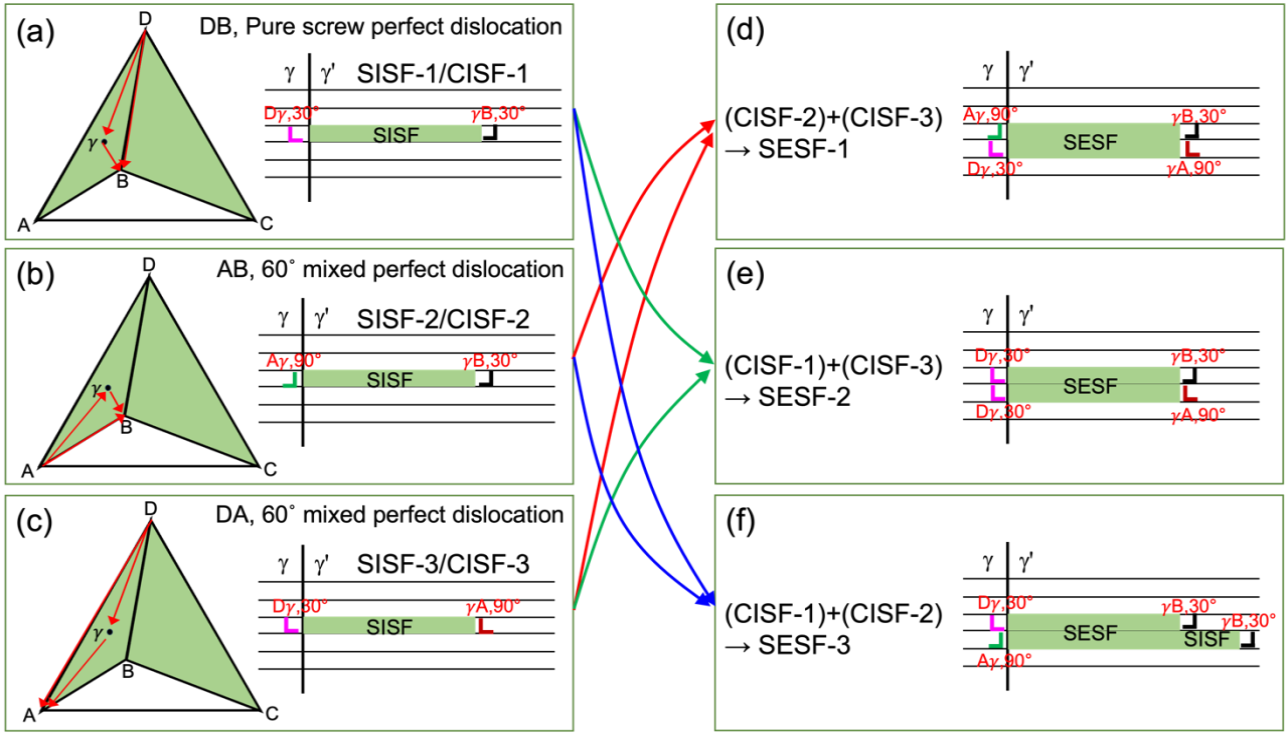

**Fig. 10** (a), (b) and (c): Schematic illustrations and Thompson tetrahedrons showing dislocation dissociation and synthetic reactions that lead to three different SISFs that SISF-1, SISF-2 and SISF-3, which have been observed directly in superalloys by Vorontsov et. al. (*Vorontsov et. al. 2012*) and Smith et. al. (*Smith et. al. 2018*), respectively. **DB** is a perfect pure screw dislocation and **AB, DA** are perfect 60° mixed dislocations. (d), (e) and (f): Schematic illustrations showing the formation of SESFs from synthetic reactions between two different CISFs of the types shown in (a), (b) and (c). To distinguish between different partial dislocations, dislocation signs were indicated using distinct colors (The more details were shown in **Fig. 4(b)**).

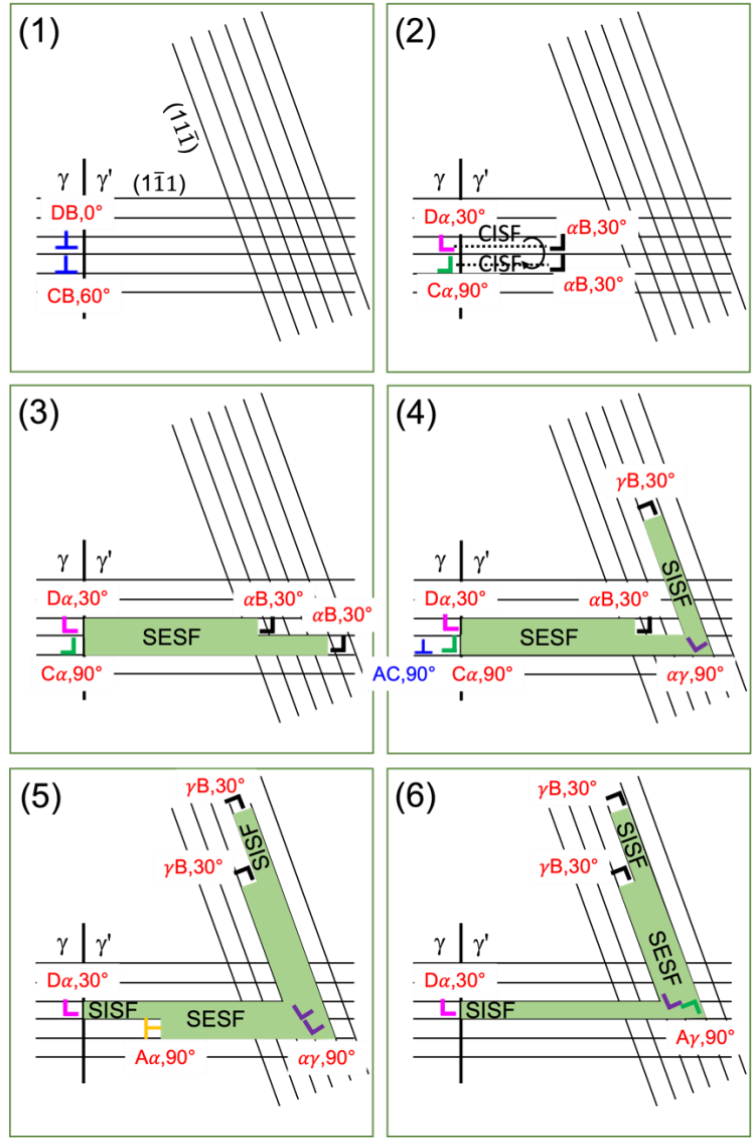

**Fig. 11.** Schematic model of L-like SF interaction via cross-slip of Shockley partial dislocation.

**Fig. 11** presents a schematic model of the L-like SF interaction corresponding to the experimentally found configuration shown in **Fig. 5**. By comparing the structure of the initial SESF with the configurations shown in **Fig. 10(f)**, it is obvious that the L-like SF interaction involves a SESF-3 mode, described in **Appendix 1**, which arises during the cutting of γ′ precipitates by two perfect $a/2\langle 110 \rangle$ dislocations: specifically, an $a/2\,[011]$ dislocation with pure screw character and an $a/2\,[\bar{1}01]$ dislocation with 60° mixed character. In the following sections, we provide a detailed description of the shear process and cross-slip mechanisms associated with the interaction shown in **Fig. 5** based on the schematic illustrations in **Figs. 10(f)** and **11**.

(1) There are two perfect dislocations (blue) gliding in two adjacent $(1\bar{1}1)$ crystal planes until they reach the of γ/γ′ interface.

(2) During precipitate shearing, the perfect pure screw dislocation **DB** = $a/2\,[011]$ and the perfect 60° mixed dislocation **CB** = $a/2\,[\bar{1}01]$ begin to dissociate into two Shockley partial dislocations, respectively. The two leading partials, **αB** = $a/6\,[\bar{1}12]$ (black), glide into the interior of the γ′ precipitate, while the trailing partials, **Dα** (pink) and **Cα** (green), remain pinned at the γ/γ′ interface. The reaction is:

$$\underset{a/2\,[\bar{1}01]}{\mathbf{CB},60°} \rightarrow \underset{a/6\,[\bar{2}\bar{1}1]}{\mathbf{C\alpha},90°} + \text{CISF} + \underset{a/6\,[\bar{1}12]}{\mathbf{\alpha B},30°}$$

$$\underset{a/2\,[011]}{\mathbf{DB},0°} \rightarrow \underset{a/6\,[121]}{\mathbf{D\alpha},30°} + \text{CISF} + \underset{a/6\,[\bar{1}12]}{\mathbf{\alpha B},30°} \quad (1)$$

(3) The dislocation **αB** = $a/6\,[\bar{1}12]$ begins to cross-slip from the primary $(1\bar{1}1)$ glide plane to the secondary $(11\bar{1})$ glide plane. As a result, the Burgers vector of the leading dislocation changes from **αB** = $a/6\,[\bar{1}12]$ to **γB** = $a/6\,[\bar{1}21]$. Meanwhile, a stair-rod edge dislocation, **αγ** = $a/6\,[0\bar{1}1]$, remains at the cross-slip point as a result of the reaction:

$$\underset{a/6\,[\bar{1}12]}{\mathbf{\alpha B},30°} \rightarrow \underset{a/6\,[0\bar{1}1]}{\mathbf{\alpha\gamma},90°} + \underset{a/6\,[\bar{1}21]}{\mathbf{\gamma B},30°}\,(\text{cross}-\text{slip}) \quad (2)$$

(4) After the two leading partials cross-slip, two sessile stair-rod dislocations **αγ** remain at the cross-slip point. At the γ/γ′ phase interface, a trailing Shockley edge partial **Cα** attracts and combines with a perfect edge dislocation (**AC** = $a/2\,[01\bar{1}]$) to form a Frank dislocation (**Aα** = $a/3\,[\bar{1}1\bar{1}]$):

$$\underset{a/2\,[01\bar{1}]}{\mathbf{AC},90°} + \underset{a/6\,[\bar{2}\bar{1}1]}{\mathbf{C\alpha},90°} \rightarrow \underset{a/3\,[\bar{1}1\bar{1}]}{\mathbf{A\alpha},90°} \quad (3)$$

(5) Subsequently, the trailing Frank partial, **Aα** = $a/3\,[\bar{1}1\bar{1}]$, originally on the primary $(1\bar{1}1)$ plane, migrates toward a sessile stair-rod dislocation (**αγ**), driven by their large junction angle ($\arccos((\mathbf{A\alpha}\cdot\mathbf{\alpha\gamma})/(\|\mathbf{A\alpha}\|\cdot\|\mathbf{\alpha\gamma}\|)) = 145°$) and mutual attraction. As **Aα** climbs, it converts the higher-energy SESF ($\gamma_{SESF} = 0.089$ J/m$^2$) into a lower-energy SISF ($\gamma_{SISF} = 0.068$ J/m$^2$) in the ordered L1$_2$ Ni$_3$Al (*Eurich et.al. 2015*), thereby reducing the system's energy. When **Aα** and **αγ** meet, they react as follows:

$$\underset{a/3\,[\bar{1}1\bar{1}]}{\mathbf{A\alpha},90°} + \underset{a/6\,[0\bar{1}1]}{\mathbf{\alpha\gamma},90°} \rightarrow \underset{a/6\,[\bar{2}1\bar{1}]}{\mathbf{A\gamma},90°} \quad (4)$$

Consequently, at the cross-slip point, there remain one Shockley edge partial **Aγ** = $a/6\,[\bar{2}1\bar{1}]$ and one stair-rod edge dislocation **αγ** = $a/6\,[0\bar{1}1]$ in the final configuration, as observed in **Fig. 5(g)**.

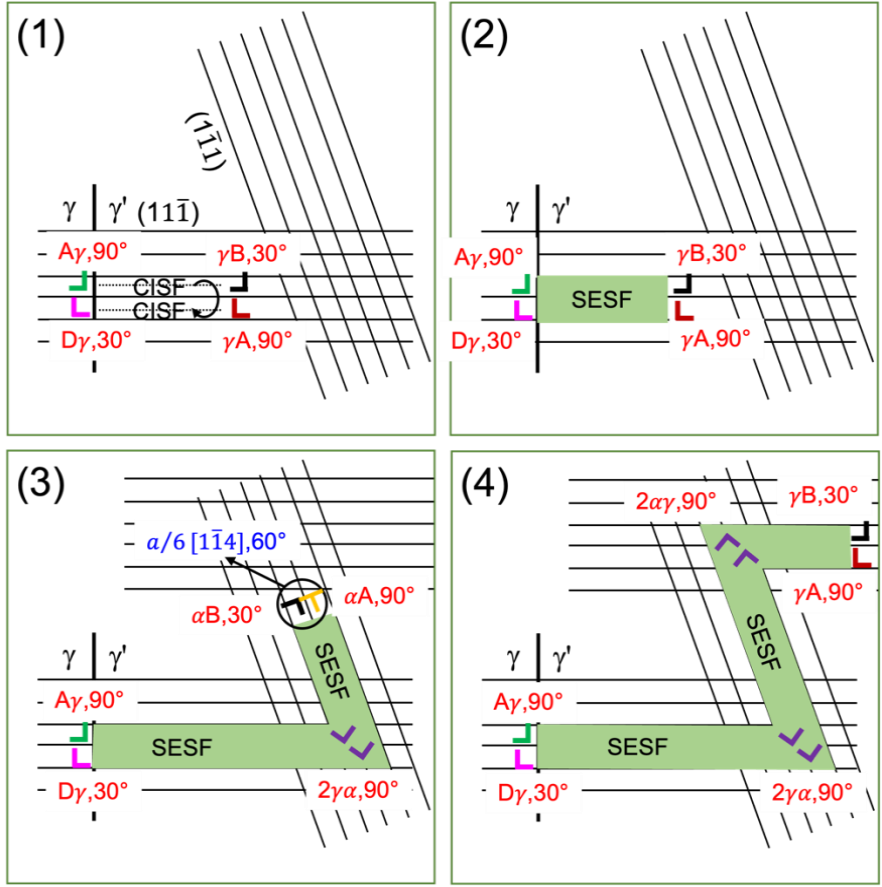

**Fig. 12.** Schematic model of Z-like SF interaction via double cross-slip of Shockley partial dislocation.

**Fig. 12** presents a schematic model of the mechanism of the Z-like SF interaction, for which an experimental example is shown in **Fig. 6**. This Z-like SF interaction is associated with the SESF shearing mode SESF-1, as described in **Fig. 10(d)** and **Fig. A.2**. There are two modes to explain the Z-like SF interaction. The first mode (mode 1) was described below in steps (1)-(4).

(1) Based on reaction Eq. (1) in **Appendix 1**, the process involves the dissociation of two 60° mixed dislocations into pairs Shockley partials which enclose a **CISF**:

$$\begin{matrix} \mathbf{DA, 60°} \\ a/2\,[101] \end{matrix} \rightarrow \begin{matrix} \mathbf{D\gamma, 30°} \\ a/6[112] \end{matrix} + \mathbf{CISF} + \begin{matrix} \mathbf{\gamma A, 90°} \\ a/6[2\bar{1}1] \end{matrix} \quad (5a)$$

$$\begin{matrix} \mathbf{AB, 60°} \\ a/2\,[\bar{1}10] \end{matrix} \rightarrow \begin{matrix} \mathbf{A\gamma, 90°} \\ a/6[\bar{2}1\bar{1}] \end{matrix} + \mathbf{CISF} + \begin{matrix} \mathbf{\gamma B, 30°} \\ a/6[\bar{1}21] \end{matrix} \quad (5b)$$

(2) Following elemental segregation, the SF type transforms from two consecutive CISFs to a single SESF through a reordering process.

(3) The leading dislocations, 90° **γA** = $a/6\,[2\bar{1}1]$ and 30° **γB** = $a/6\,[\bar{1}21]$, initiate *cross-split* and cross-slip from the primary $(11\bar{1})$ glide plane to the secondary $(1\bar{1}1)$ glide plane. As analyzed in **Fig. 6(h)**, two stair-rod dislocations, **2γα** = $a/3\,[01\bar{1}]$, remain at the primary *cross-split* / cross-slip point through the following reaction:

$$\underset{a/6[2\bar{1}1]}{\gamma A, 90°} \rightarrow \underset{a/6[01\bar{1}]}{\gamma \alpha, 90°} + \underset{a/3[1\bar{1}1]}{\alpha A, 90°} \text{ (cross – split)} \tag{6a}$$

$$\underset{a/6[\bar{1}21]}{\gamma B, 30°} \rightarrow \underset{a/6[01\bar{1}]}{\gamma \alpha, 90°} + \underset{a/6[\bar{1}12]}{\alpha B, 30°} \text{ (cross – slip)} \tag{6b}$$

The 90° edge dislocations **γA** can be split into two other 90° edge dislocations (**γα** and **αA**) that glide on two different crystallographic planes which is called "*cross-split* of edge dislocations" (*Kositski et.al. 2016*).

(4) Subsequently, the leading dislocations, a 90° Frank partial **αA** = $a/3\,[1\bar{1}1]$ and a 30° Schockley partial **αB** = $a/6\,[\bar{1}12]$, undergo secondary *cross-split* and cross-slip, returning from the secondary $(1\bar{1}1)$ glide plane to the primary $(11\bar{1})$ glide plane. As a result of this process, two stair-rod edge dislocations, **2αγ** = $a/3\,[0\bar{1}1]$, remain at the secondary *cross-split* / cross-slip point which have formed through the following reaction:

$$\underset{a/3[1\bar{1}1]}{\alpha A, 90°} \rightarrow \underset{a/6[0\bar{1}1]}{\alpha \gamma, 90°} + \underset{a/6[2\bar{1}1]}{\gamma A, 90°} \text{(cross – split)} \tag{7a}$$

$$\underset{a/6[\bar{1}12]}{\alpha B, 30°} \rightarrow \underset{a/6[0\bar{1}1]}{\alpha \gamma, 90°} + \underset{a/6[\bar{1}21]}{\gamma B, 30°} \text{ (cross – slip)} \tag{7b}$$

(5) Alternatively, there is another mode (mode 2). Because of the attractive interactions between **Dγ** (pink) and **Aγ** (green), and between **γA** (red) and **γB** (black), reaction Eqs. (5) may alternatively be written as:

$$\underset{a/2[101]}{DA, 60°} + \underset{a/2[\bar{1}10]}{AB, 60°} \rightarrow \underset{a/2[011]}{DB, 0°} \rightarrow \underset{a/6[\bar{1}21]}{\gamma B, 30°} + \text{SESF} + \underset{a/6[112]}{D\gamma, 30°} \tag{8}$$

The leading partial dislocation, **Dγ** = $a/6\,[112]$, glides into the interior of the γ′ precipitate, while the trailing partial, **γB** = $a/6\,[\bar{1}21]$, remains pinned at the γ/γ′ phase interface. The leading dislocation **Dγ** = $a/6\,[112]$ initiates cross-slip from the primary $(11\bar{1})$ glide plane to the secondary $(1\bar{1}1)$ glide plane. The two stair-rod dislocations, **2γα** = $a/3\,[01\bar{1}]$, are generated at the primary cross-slip point through the following reaction:

$$\underset{a/6[112]}{D\gamma, 30°} \rightarrow \underset{a/3[01\bar{1}]}{2\gamma \alpha, 90°} + \underset{a/6[1\bar{1}4]}{\frac{D\gamma}{2\gamma \alpha}, 60°} \text{ (cross – slip)} \tag{9}$$

(6) The **Dγ/2γα** = $a/6\,[1\bar{1}4]$ can be the product of the synthetic reaction between a 90° Frank partial **αA** = $a/3\,[1\bar{1}1]$ and a 30° Schockley partial **αB** = $a/6\,[\bar{1}12]$. Subsequently, the leading dislocation **Dγ/2γα** = $a/6\,[1\bar{1}4]$ dissociates again and one part of it undergoes secondary cross-slip, returning from the secondary $(1\bar{1}1)$ glide plane to the primary $(11\bar{1})$ glide plane. During this process, the leading dislocation Burgers vector changes from **Dγ/2γα** = $a/6\,[1\bar{1}4]$ back to **Dγ** = $a/6$ [112]. Meanwhile, two stair-rod edge dislocations, $2αγ = a/3$ $[0\bar{1}1]$, remain at the secondary cross-slip point through the following reaction:

$$\underset{a/6[1\bar{1}4]}{\frac{D\gamma}{2\gamma\alpha}, 60°} \rightarrow \underset{a/3[0\bar{1}1]}{2\alpha\gamma, 90°} + \underset{a/6[112]}{D\gamma, 30°} (\text{cross}-\text{slip}) \quad (10)$$

Based on the L-like and Z-like stacking fault interactions described above, the cross-slip of a 30° Shockley partial or the *cross-split* of a 90° Shockley partial within γ′ precipitates in superalloys can be activated through the formation of stair-rod dislocations at the intersection line of two conjugate {111} planes.

### 4.2. Element segregation effect on defect energy

The above mechanism involves complex dislocation reactions. In some cases, these reactions require overcoming high energy barriers, which may hinder their occurrence. However, elemental segregation can decrease these energy barrrier, inlcuding SF energy and dislocation line energy. Here, we analyze the dislocation reaction process from an energetic perspective. The relationship between solute segregation and segregation energy ($\Delta H_{seg}$) typically follows the Langmuir-Mclean isotherm (*Langmuir 1918, McLean 1957, Murdoch et.al. 2013, Malik et.al. 2020*), which follows as:

$$\frac{X_{int}}{1-X_{int}} = \frac{X_{bulk}}{1-X_{bulk}} exp\left(\frac{-\Delta H_{seg}}{kT}\right) \quad (11)$$

where $k$ is boltzmann constant, $T$ is temperature, $X_{int}$ is mole fraction of the interface (indicates SFs in this work) and $X_{bulk}$ is mole fraction of the γ′ phase. The solute excess Γ at the interface can be expressed via (*Trelewicz, 2009*):

$$\Gamma_{int} = \frac{(X_{int}-X_{bulk})\cdot d}{\Omega\cdot(1-X_{bulk})} \quad (12)$$

where $\Omega$ is atom volume in L1$_2$ structure Ni$_3$Al, $\Omega \cdot (1 - X_{bulk}) \cdot N_A$ is mole volume of the solvent, $N_A$ is Avogadro constant (6.02214076×10²³ mol⁻¹) and $d = \frac{d2-d1}{2}$ is the thickness of interface, where $d_1$ and $d_2$ are determined by the line profile of STEM-EDS, see **supplementary materials 1 Fig. S2**.

When the segregation reaches saturation but not make new phase formation, the defect energy change (indicates SF energy in this work) can be expressed (c.f. Fig. 1 in reference (*Kirchheim 2007*)) in binary alloys:

$$\Delta\gamma_{int} = -\Gamma_{int} \cdot \Delta H_{seg} \quad (13)$$

For a multicomponent alloy, the defect energy change can be expressed via:

$$\Delta\gamma_{int} = -\sum_i^n X_{bulk}^i \cdot \Gamma_{int}^i \cdot \Delta H_{seg}^i \quad (14)$$

In this work, the stair-rod dislocation can be regarded as a 1D interface. The solute excess Γ at a single dislocation is the additional amount of solute segregated at a dislocation compared to the bulk matrix, which can be expressed as:

$$\Gamma_{dis} = \frac{N_{solute}}{L_{dis}} \tag{15}$$

where $N_{solute}$ is the total number of solute atoms segregated at the dislocation and $L_{dis}$ the length of the dislocation in the analyzed region. From STEM-EDS mapping, we can determine solute excess per unit dislocation:

$$\Gamma_{dis} = \frac{(X_{dis}-X_{bulk}) \cdot A}{\Omega(1-X_{bulk})} \tag{16}$$

where $X_{dis}$ is mole fraction at the dislocation, $A = \pi r^2$ is the integration area (cross-section normal to the dislocation line) (*Kirchheim 2007*) and $r$ denotes core radius and the size of the strain field of the dislocation, which can be determined based on elements distribution determined by STEM-EDS map, see **supplementary materials 1 Figs. 3**. At a grain boundary, the defects are usually dislocation arrays (*Medouni et.al. 2021*), therefore, the segregation energy at dislocations can be also calculated by Eq. (11). The defect energy change at a dislocation of a multicomponent alloy should be expressed via:

$$\Delta U_{dis} = -\sum_i^n X_{bulk}^i \cdot \Gamma_{dis}^i \cdot \Delta H_{seg}^i \tag{17}$$

Based on Eq. (14), we can have SISF energy reduction in a L-like SF interaction, $\Delta\gamma_{SISF} = -0.058 \, J/m^2$ and SESF energy reduction in a Z-like SF interaction, $\Delta\gamma_{SESF} = -0.071 \, J/m^2$, respectively. Based on Eq. (17), we can calculate the dislocations line energy reduction in a L-like SF interaction, $\Delta U_{stair-L} = -0.67 \times 10^{-10} J/m$ and dislocations line energy reduction in a Z-like SF interaction, $\Delta U_{stair-Z} = -0.63 \times 10^{-10} J/m$, respectively. The calculations was done in excel, which was shown in **supplementary materials 2**.

### 4.3. The activation energy of Shockley partial dislocation cross-slip

For the Fleischer mechanism, the stacking fault extends into the cross-slip plane without being completely constricted to a point on the glide plane. The saddle configuration contains partial dislocations on both the glide and the cross-slip planes, as well as a stair-rod dislocation at their intersection. One stair-rod edge dislocation $\alpha\gamma = a/6\,[0\bar{1}1]$ remains in the cross-slip point by reaction:

$$\begin{array}{c} \alpha B, 30° \\ a/6\,[\bar{1}12] \end{array} \rightarrow \begin{array}{c} \alpha\gamma, 90° \\ a/6\,[0\bar{1}1] \end{array} + \begin{array}{c} \gamma B, 30° \\ a/6\,[\bar{1}21] \end{array} \tag{18}$$

where $|\mathbf{b}_{\alpha B}|^2 < |\mathbf{b}_{\alpha \gamma}|^2 + |\mathbf{b}_{\gamma B}|^2$ indicates this reaction does not occur spontaneously and this reaction configuration is unstable. This reaction needs external activation energy done by large applied stress. Cai et al. indicated that Fleischer mechanism dominates in the high-stress limit (*Kuykendall et.al. 2020*). According to the angle of SFs interaction, we can have two Fleischer modes of cross-slip: acute angle mode (70.53°) and obtuse angle mode (109.47°). In this work, we provide acute angle mode in **Appendix 2 (case 1)**. In addition, we summarised the relationship of activation energy, stacking fault energy, dislocation interaction energy, dislocation self-energy and applied stress in cross-slip plane. The activation energy of acute angle mode for Fleischer mechanism can be expressed as:

$$\Delta E = \frac{K_e b^2}{3}\left[\frac{1}{6}\ln\frac{r_{34}}{r_4} - \frac{1}{12}\ln\frac{r_0}{r_{13}}\right] + \frac{K_e b^2}{108}[\sqrt{2}\sin\alpha + 3\sqrt{2}\sin2\theta + 5\cos2\alpha - \frac{3}{2}\cos2\theta + \frac{9}{2}]$$

$$+ \frac{K_s b^2}{4}\ln\frac{r_0}{r_{13}} - \left(\frac{\tau b}{2} - \gamma_{SISF}\right)r_{34} \quad (19)$$

where $K_e = G/2\pi(1-\nu)$, $K_s = G/2\pi$, $\nu = 0.28$ is Poisson's ratio, $\tau$ is the applied shear stress in cross-slip plane, $\gamma_{SISF}$ is the SISF energy, $G$ is shear modulus of the γ′ phase at 1123 K, which is calculated based on $G = 78.86 - 0.0009 \times T - 0.000019 \times T^2$ (GPa) (*Galindo-Nava et.al. 2015*). $b$ is the magnitude of Burgers vector $a/2[011]$ and $r_4 = 2b/3$. (The physical meaning of the other signs, such as $r_{34}$, $r_0$ $r_{13}$, $\alpha$ and $\theta$, was described in **Appendix 2 (case 1)** ). The maximum of $\Delta E$ ($\Delta E_m$) can be regarded as the activation energy for acute cross-slip. According to the diagram in **Fig. A3**, we can have

$$r_{13} = \sqrt{(r_0 \sin\alpha)^2 + (r_{34} - r_0\cos\alpha)^2} \quad (20)$$

$$\theta = \arctan\left(\frac{r_{34}\sin\alpha}{r_0 - r_{34}\cos\alpha}\right) \quad (21)$$

where $\alpha = 70.53°$ is the acute angle formed between the primary and cross-slip planes, $r_0 = 102\ nm$ is the distance of trailing dislocation at the interface of γ/γ′ phases and the first cross-slip point in this work, based on HAADF-STEM image in **Figs. 5 (a)** and **(b)**. Then, the $\Delta E_m$ is numerically computed via numerical analysis. The relationship of applied shear stress and activation energy is shown in **Fig. 13(a)**. Based on Eq. (19), the activation energy strongly depends on applied shear stress, dislocation line energy and SF energy. It was known that element segregation around SFs will reduce the formation energy of SFs (*Smith et.al. 2015, Rao et.al. 2018*). For instance, the SISF energy in CoNi-based superalloys will decrease from $\gamma_{SISF} = 0.08\ J/m^2$ down to $\gamma_{SISF} = 0.02\ J/m^2$ (*Rao et.al.*

*2018*). According to the result in **Fig. 13(a)**, the decrease of SF energy facilitates an activation energy reduction of cross-slip. In addition, the element segregation around dislocations can also decrease their line energy by about 33% (*Kirchheim 2007*). Therefore, we can put element segregation effect on SF energy and dislocation line energy in Eq. (20) by introducing two additional coefficients:

$$\Delta E_c = \frac{K_e b^2}{3}\left[\frac{1}{6}\eta_1 \ln\frac{r_{34}}{r_4} - \frac{1}{12}\ln\frac{r_0}{r_{13}}\right] + \frac{K_e b^2}{108}[\sqrt{2}\sin\alpha + 3\sqrt{2}\sin2\theta + 5\cos2\alpha - \frac{3}{2}\cos2\theta + \frac{9}{2}]$$

$$+ \frac{K_s b^2}{4}\ln\frac{r_0}{r_{13}} - \left(\frac{\tau b}{2} - \eta_2 \gamma_{SISF}\right)r_{34} \qquad (22)$$

where $\eta_1 = 0.65$ is coefficient representing the element segregation effect on the elastic energy of the stair-rod dislocation and $\eta_2 = 0.15$ is coefficient representing the element segregation effect on the SF energy. In this work, according to calculations in **section 4.2**, the SISF energy reduction caused by elements segregation in L-like SF interaction is $\Delta\gamma_{SISF} = -0.058$ J/m², about 85% reduction compared with $\gamma_{SISF} = 0.068$ J/m² of L1$_2$ structured Ni$_3$Al (*Eurich et.al. 2015*). $\Delta U_{stair-L} = -0.67 \times 10^{-10}$ J/m is the line energy reduction of a stair-rod dislocation ($\boldsymbol{b} = \frac{a}{6}[01\bar{1}]$) in a L-like SF interaction being lowering by about 35% due to element segregation, compared with line energy $U = \alpha G b^2$, where $G$ is taken to be 53.9 GPa at 850 °C. Therefore, element segregation around the stair-rod dislocation and at the SFs will decrease the activation energy which is beneficial for dislocation cross-slip via stair-rod formation. However, even with considering these segregation effects the activation of cross-slip still needs quite high applied shear stress.

The cross-slip / *cross-split* frequency is related to the activation energy $\Delta E_c$ and temperature (*Li et.al. 2021*), which can be expressed by:

$$f = \omega L/L_0 \exp(\Delta E_0/k_B T) \qquad (23)$$

where $\omega = 10^{12} s^{-1}$ is the Debye frequency for strain rate $\dot{\varepsilon} = 10^{-4} s^{-1}$, $L = 176\ nm$ is dislocation length equal to the diameter of $\gamma'$ particle, $L_0 = 1\ um$ is the reference length as defined in ref. (*Hussein et.al. 2015*), $T$ is 1123 K.

The cross-slip / *cross-split* probability in a certain period of time:

$$p(t) = 1 - \exp(-f \cdot t) \qquad (24)$$

where $t = 340\ s$ is the whole time during alloy compression test. Because the investigated TEM sample experienced 340 s during deformation test. Based on Eqs. (23) and (24), the relationship of cross-slip / *cross-split* probability and activation energy is shown in **Fig. 13(b)**. When activation energy is higher than a critical value of 3.6 eV, the cross-slip / *cross-split* probability is zero.

Therefore, we can get the required minimum critical stress for 30° partial cross-slip and 90 partial *cross-split*, as shown in **Table 4**.

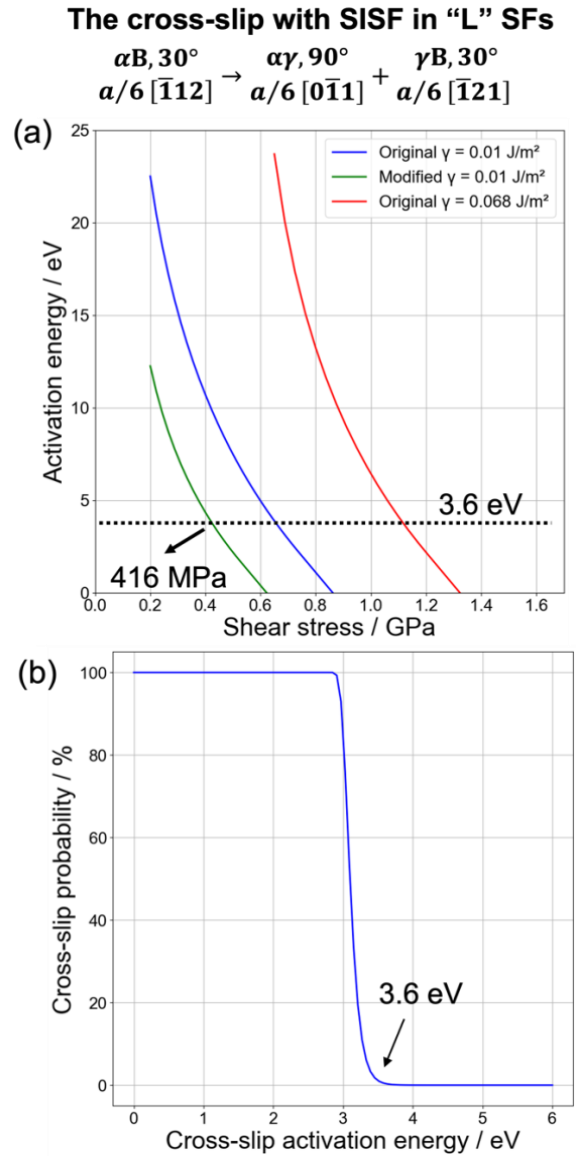

**Fig. 13.** (a) Relationship of 30º Shockley partial cross-slip activation energy and required shear stress based on Eqs. (19) and (22). Element segregation reduces the stair-rod dislocation line energy and SISF energy, which reduces the activation energy and necessary shear stress for cross-slip. (**Red line:** there is no any segregation effect. **Blue line:** the element segregation makes SF energy reduction. **Green line:** the element segregation makes SF energy and dislocation line energy reduction) (b) The relationship of cross-slip / *cross-split* probability and activation energy. When activation energy is higher than 3.6 eV, the cross-slip / *cross-split* probability is zero.

**Table 4.** Minimum critical stress required to reach an activation energy of 3.6 eV, as determined from reaction (2) of the L-like stacking fault (SF) interaction configuration ("L"), and from reactions (6), (7), (9), and (10) of the Z-like SF interaction configuration ("Z").

|  | "L" with reaction (2) | "Z" with reactions (6) and (7) (mode 1) | | "Z" with reactions (9) and (10) (mode 2) | |
| --- | --- | --- | --- | --- | --- |
|  |  | The first | The second | The first | The second |
| Cross-slip | 416 MPa | 421 MPa | 420 MPa | *1024 MPa* | 154 MPa |
| Cross-split | - | 449 MPa | 99 MPa | - | - |

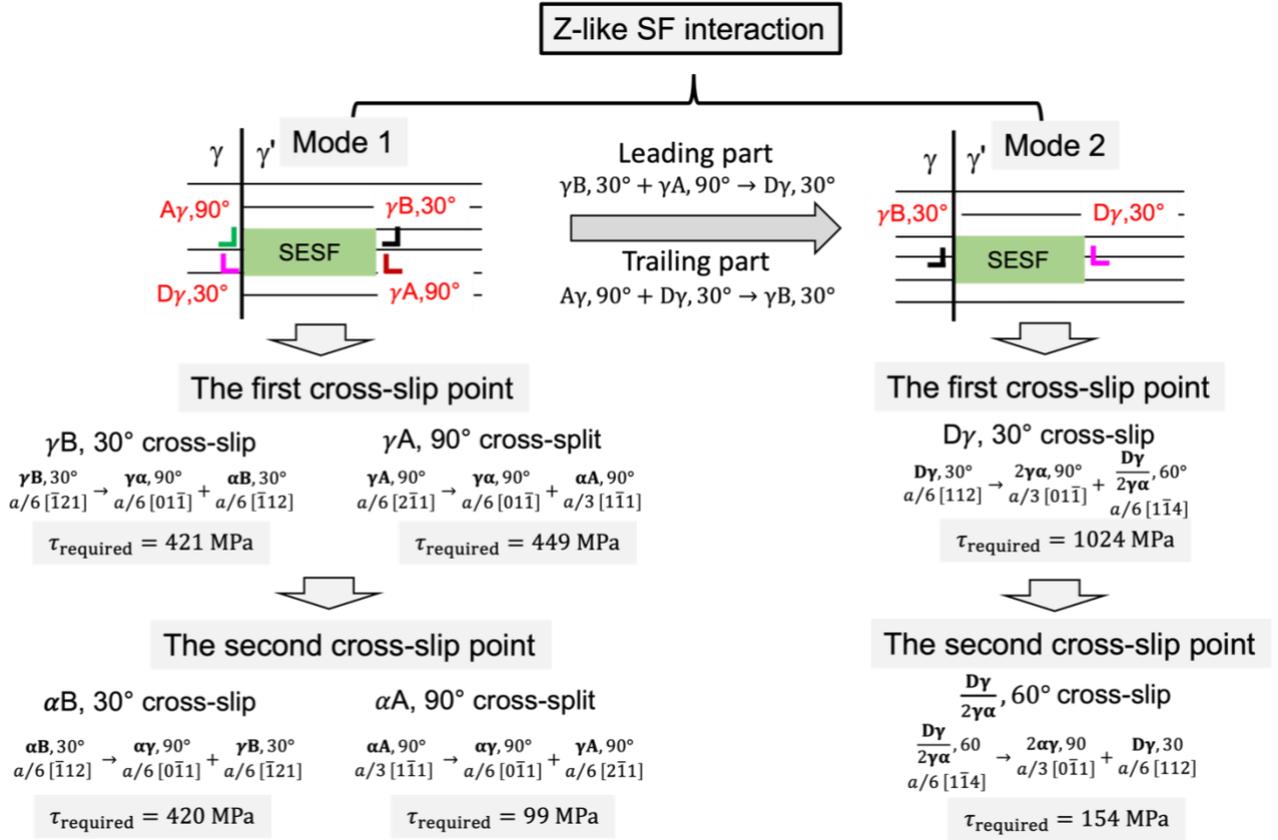

**Fig. 14.** Schematic summary of the two modes for explaining the Z-like SF interaction: **Mode 1** and **Mode 2**.

To explain the Z-like SF interaction, two distinct mechanisms have been proposed, see **Fig. 14**. **Mode 1** is based on reaction Eqs. (6) and (7). In this mode, the leading segment consists of two dislocations: a 90º Shockley partial (γA) and a 30º Shockley partial (γB). While the 90º edge dislocation cannot, in principle, cross-slip into other slip planes, it can undergo a *cross-split* process, decomposing into a stair-rod dislocation (γα) and a 90º Frank partial (αA). The mathematical model describing this *cross-split* of edge dislocations is provided in **Appendix 2 (Case 2)**. According to our model, the critical resolved shear stress is required to reach the minimum applied stress at which the activation energy for cross-slip becomes 3.6 eV. The calculated critical resolved shear stresses for the first cross-slip (421 MPa) and the first *cross-split* (449 MPa) are comparable, as summarized in **Table 4**.

**Mode 2**, in contrast, is based on reaction Eqs. (9) and (10). In this case, the leading dislocation is a 30º Shockley partial (**Dγ**), where **Dγ** = **γA** + **γB**. However, experimental observations reveal that the first cross-slip point involves two stair-rod dislocations (2**γα**). The resulting cross-slip dislocation is a 60º Shockley partial ($\frac{D\gamma}{2\gamma\alpha} = a/6\,[1\bar{1}4]$), which has been previously observed in other fcc systems by TEM investigation (*Geipel et.al. 1993*). Nonetheless, this reaction significantly increases the dislocation line energy, as indicated by the inequality $|\mathbf{b_{D\gamma}}|^2 \ll |\mathbf{b_{2\gamma\alpha}}|^2 + |\mathbf{b}_{a/6[1\bar{1}4]}|^2$. Consequently, this reaction demands a much higher minimum critical stress of 1024 MPa, more than twice that required for the first cross-slip or *cross-split* in **Mode 1**, as shown in **Table 4**. The corresponding mathematical model for the cross-slip of a 30º Shockley partial involving two stair-rod dislocations is described in detailed in **Appendix 2 (Case 3)**.

**Mode 1** requires lower activation energy and stress compared to **Mode 2**. Therefore, **Mode 1**, which involves a 90º edge dislocation *cross-split*, is a more suitable model to explain the Z-like SF interaction.

**4.4. Local stress concentration in the γ′ phase**

Resolved shear stress (RSS) is related to the applied stress by a geometrical factor, *m*, typically called the Schmid factor:

$$\tau_{RSS} = \sigma_{app} m = \sigma_{app}(cos\phi cos\lambda) \qquad (25)$$

where $\sigma_{app}$ is the magnitude of the applied tensile stress, $\phi$ is the angle between the normal of the slip plane and the direction of the applied force, and $\lambda$ is the angle between the slip direction and the direction of the applied force. The maximum value of geometrical factor *m* is $cos\phi cos\lambda$ = 0.5. In this work, the maximum applied stress is 700 MPa at strain 3.4 %, according to stress-strain curve in **Fig. 2**. Therefore, the maximum resolved shear stress is $\tau_{RSS}^{max} = \sigma_{RSS}^{max} 0.5 = 350$ MPa, which is lower than the required shear stress (416 MPa) for dislocation cross-slip with SISF in the L-like SF interaction and (450 MPa) for dislocation *cross-split* with SISF in Z-like SF interaction. Therefore, a high local stress concentration is necessary to explain the activation of dislocation cross-slip. When a certain of plastic strain is reached, dislocations approach each other and pile up which will induce quite high local stress around the γ′ particles. By analysis, beside of the applied stress $\sigma_{app}$, the local stress concentration stems mainly from dislocation pile-ups $\sigma_{dis}$ at the interface and shearing dislocations in the γ′ precipitates, misfit stress $\sigma_{misfit}$ at the interface between the γ and γ′ phases and dislocation itself line tension $\sigma_{self}$ (0 for straight dislocation with infinite length):

$$\sigma_{total} = \sigma_{app} + \sigma_{misfit} + \sigma_{dis} + \sigma_{self} \tag{26}$$

In CoNi-based superalloys, the positive misfit between the γ and γ′ phases is very small when temperature is close to 900 °C (*Xue et.al. 2018*). Accordingly, the contribution of misfit stress is very small at high temperature which can be neglected. The stress $\sigma_{dis}$ contributed from surrounding dislocations can be expressed by:

$$\sigma_{dis} = \sigma_{dis}^{sh} + \sigma_{dis}^{int} = \sum_{0}^{n_{sh}} \frac{Gb_{sh}}{2\pi r_{sh}} + \sum_{0}^{n\_int} \frac{Gb_{int}}{2\pi r_{int}} \tag{27}$$

where *n_sh* is the shearing dislocation number, *n_int* is the shearing dislocation number, $r_{sh}$ is the distance between cross-slip point and other shearing dislocations in the γ′ precipitates, $r_{int}$ is the distance between cross-slip point and dislocations in the interface, $b_{sh}$ is the the magnitude of Burgers vector of the shearing dislocations, $b_{int}$ is the the magnitude of Burgers vector of the dislocations (pile-ups) in the interface (see **supplementary materials 1 Fig. S4**) and *G* is the shear modulus, 53.9 GPa at 850 °C.

In the case of the L-like SF interaction, there is one shearing dislocation marked with SF1 which has a distance $r_{sh}$ =16 nm from the cross-slip point, shown in **supplementary materials 1 Fig. S5(a)**. According to Burgers circuit, shown in **supplementary materials 1 Fig. S5(b)**, the Burgers vector is deduced to be $a/3\,[\bar{1}1\bar{1}]$, identifying it as a Frank edge dislocation. Therefore, $b_{sh}$ = 0.21 nm and $\sigma_{dis}^{sh}$ can achieve 155 MPa which stress contribution is not enough high for cross-slip activation. The average distance of the first cross-slip point and all perfect dislocations at the γ′ particle interface can be assumed to be equal to the radius of the γ′ precipitate, $r_{int}$ = 118 nm. In addition, $b_{int}$ = 0.25 nm for perfect dislocations $a/2\,\langle 011 \rangle$ dislocations and *n_int* = 12 (see **supplementary materials 1 Fig. S4**) and $\sigma_{dis}^{int}$ can reach 219 MPa at the cross-slip point. The total local stress $\sigma_{total} = \sigma_{app} + \sigma_{dis}$ = 1074 MPa and therefore the maximum resolved shear stress is $\tau_{RSS}^{max} = \sigma_{total} \cdot 0.5$ = 537 MPa, which is higher than the required shear stress (416 MPa) for 30° dislocation cross-slip.

In the case of the Z-like SF interaction, there is no other shearing dislocations in the γ′ precipitate, as shown in **supplementary materials 1 Fig. S5(c)**. The first cross-slip point has an average distance $r_{int}$ = 117 nm from the perfect dislocations at the γ′ precipitate interface. Therefore, $b_{int}$ = 0.25 nm for perfect $a/2\,\langle 011 \rangle$ dislocations and *n_int* = 12 (see **supplementary materials 1 Fig. S4**) and $\sigma_{dis}^{int}$ can achieve 219 MPa which interface dislocation pile-ups makes higher contribution for cross-slip activation than limited shearing dislocations. The total local stress $\sigma_{total} = \sigma_{app} + \sigma_{dis}$ = 920 MPa and therefore the maximum resolved shear stress is $\tau_{RSS}^{max} = \sigma_{total} \cdot 0.5$ = 460 MPa, which is higher than the required shear stress for the first dislocation cross-slip (421 MPa) and *cross-split* (450 MPa).

The second cross-slip point is at the interface and therefore much closer to interfacial dislocation pile-ups. The local stress should be higher than in the interrior of the γ′ precipitate which can easily facilitate the dislocation cross-slip / *cross-split* near the two phases interface.

Based on the discussion above, the local stress concentration is another important factor to trigger Schockley partials cross-slip / *cross-split* in addition to the defects energy reduction with elements segregation.

## 5. Summary

In this work, we discovered two new types of SF interactions in the L1$_2$-γ′ phase of CoNi-base superalloys, i.e., the L-like SF interaction and the Z-like SF interaction. Their formation is associated with the cross-slip or *cross-split* of Shockley partial dislocations through the formation of stair-rod dislocations within the γ' phase of superalloys. This finding further supports the previous assertion that the V-like SF interaction is generated by cross-slip or *cross-split* of Shockley partial dislocations, being expected to occur extensively in the superalloys. Despite these findings, the existence of a Lomer-Cottrell assisted nucleation mechanism for V-like SF interaction cannot be ruled out.

In addition, we proposed a simple mathematical model to explain this phenomenon. The activation energy for cross-slip or *cross-split* of Shockley partial dislocations is strongly influenced by the dislocation line energy, the SF energy, and the local stress. Elemental segregation can promote cross-slip or *cross-split* by reducing the activation energy required, primarily by lowering defect energies such as SF energy and stair-rod dislocation line energy. In addition to elemental segregation, local stress concentration, resulting from the combined effects of applied stress, shearing dislocations within the γ' phase, or dislocation pile-ups at the γ/γ′ interface, is another critical factor in triggering 30° Shockley partial cross-slip and 90° Shockley partial *cross-split*.

Moreover, the Shockley partial dislocation cross-slip / *cross-split* promotes the formation of sessile stair-rod dislocations, which is beneficial for enhancing the high-temperature deformation resistance of the superalloy.

**Appendix 1: Summary of superlattice extrinsic stacking fault (SESF) shearing modes.**

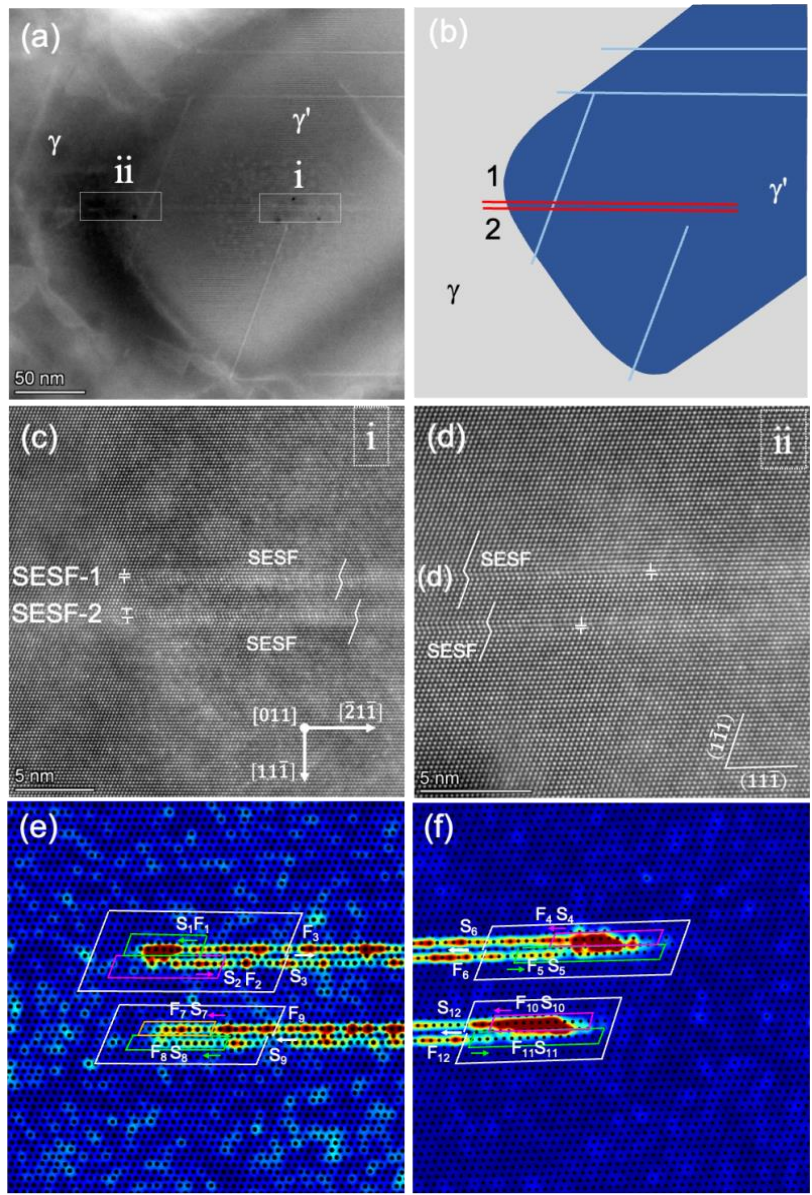

**Fig. A.1.** (a) HAADF-STEM image of γ′ particle with single SFs. (b) Schematic sketch showing the arrangement of SFs in **Fig. A.1(a)**. (c) Atomic resolution HAADF-STEM image with defects of trailing dislocations and SESFs. (d) Atomic resolution HAADF-STEM image with defects of leading dislocations and SESFs. (e) and (f) COS images for **Fig. A.1(c)** and **(d)**, respectively.

**Fig. A.1(a)** shows a single γ′ precipitates which includes several single SF configurations, marked in **Fig. A.1(b)**. By employing HRSTEM, two different SESF modes, marked as SESF1 and SESF2 in the habit plane $(11\bar{1})$ were analysed and are discussed in detail, as shown in **Figs. A.1(c) and (d)**. It is found that the leading edges as well as the trailing edges have 2 partials, respectively. And the dislocation lines of the bounding partials are all oriented along the electron beam direction $\boldsymbol{\mu} = [011]$. Clockwise-drawn Burgers circuits around the leading and trailing edges were conducted, respectively, as shown in **Fig. A.1(e) and (f)**.

In the mode **SESF-1**, it has two trailing partials, shown in **Fig. A.1(e)**, which have Burgers vectors $b_{1,p} = F_1S_1 = a/12[2\bar{1}1]$ and $b_{2,p} = F_2S_2 = a/6[\bar{2}1\bar{1}]$ in the projection plane, respectively. The real Burgers vector of the 2 trailing partials are $b_1 = a/6[112]$ (30° **Dγ** in Thompson tetrahedron) and $b_2 = a/6[\bar{2}1\bar{1}]$ (90° **γA**), respectively. The whole Burgers circuit around the trailing part was analysed and resulted in a projected Burgers vector of $b_{3,p} = F_3S_3 = a/12[\bar{2}1\bar{1}]$, which stems from a Burgers vector $b_3 = a/6[\bar{1}21]$ (30° **γA**).

Similarly, the 2 leading partials at the **SESF-1**, shown in **Fig. A.1(f)**, have Burgers vectors $b_{4,p} = F_4S_4 = a/6[2\bar{1}1]$ and $b_{5,p} = F_5S_5 = a/12[2\bar{1}1]$ in the projection plane, respectively. The real Burgers vectors of the 2 trailing partials are $b_4 = a/6[2\bar{1}1]$ (90° **γA**) and $b_5 = a/6[\bar{1}21]$ (30° **γB**), respectively. The whole Burgers circuit around the trailing part was analysed to have a projected Burgers vector $b_{6,p} = F_6S_6 = a/12[2\bar{1}1]$, which real Burgers vector is regarded as $b_6 = a/6[112]$ (30° **Dγ**).

Based on analysis, the reaction is two 60 mixed dislocations disassociation with two **CISFs**:

$$\underset{a/2\,[101]}{\mathbf{DA,60°}} \rightarrow \underset{a/6\,[112]}{\mathbf{D\gamma,30°}} + \text{CISF} + \underset{a/6\,[2\bar{1}1]}{\mathbf{\gamma A,90°}} \quad \text{(A.1.a)}$$

$$\underset{a/2\,[\bar{1}10]}{\mathbf{AB,60°}} \rightarrow \underset{a/6\,[\bar{2}1\bar{1}]}{\mathbf{A\gamma,90°}} + \text{CISF} + \underset{a/6\,[\bar{1}21]}{\mathbf{\gamma B,30°}} \quad \text{(A.1.b)}$$

where **DA** and **AB** are 60° mixed perfect dislocations, **Dγ** and **γB** are 30° Shockley partial dislocations, **Aγ** and **γA** is 90° Shockley edge partial dislocation. For the trailing partials, the angle of Shockley partials **Dγ** and **Aγ** is **120°** and therefore the trailing partials will attract from each other to become a **γB**. For the leading partials, the angle of Shockley partials **γA** and **γB** is 120° and therefore the leading partials will also attract from each other to become **Dγ**. Therefore, the reaction Eqs. (A.1) can be also written as:

$$\underset{a/2\,[101]}{\mathbf{DA,60°}} + \underset{a/2\,[\bar{1}10]}{\mathbf{AB,60°}} \rightarrow \underset{a/2\,[011]}{\mathbf{DB,0°}} \rightarrow \underset{a/6\,[\bar{1}21]}{\mathbf{\gamma B,30°}} + \text{SESF} + \underset{a/6\,[112]}{\mathbf{D\gamma,30°}} \quad \text{(A.2)}$$

where **DB** is a perfect pure screw dislocation, **Dγ** and **γB** are 30° Shockley partial dislocations.

In the mode **SESF-2**, it consists of two trailing partials and two leading partials, as shown in **Fig. A.1(e)**. The two trailing partials have projected Burgers vectors $b_{7,p} = F_7S_7 = a/12[2\bar{1}1]$ and $b_{8,p} = F_8S_8 = a/12[2\bar{1}1]$ in the projection plane, respectively. The real Burgers vectors of the two trailings are $b_7 = b_8 = a/6[112]$ (30° **Dγ** in Thompson tetrahedron). The whole projected Burgers circuit around the trailing part was analysed $b_{9,p} = F_9S_9 = a/6[2\bar{1}1]$, which real Burgers vector is most probably the one of a super partial dislocation $b_9 = a/3[112]$ (30° **2Dγ**).

The two leading partials of the **SESF-2**, shown in **Fig. A.1(f)**, have projected Burgers vectors $\mathbf{b}_{10,p} = \mathbf{F}_{10}\mathbf{S}_{10} = a/6[2\bar{1}1]$ and $\mathbf{b}_{11,p} = \mathbf{F}_{11}\mathbf{S}_{11} = a/12[\bar{2}1\bar{1}]$ in the projection plane, respectively. The real Burgers vectors of the two trailings are $\mathbf{b}_{10} = a/6[2\bar{1}1]$ (90° **γA**) and $\mathbf{b}_{11} = a/6[\bar{1}21]$ (30° **γB**), respectively. The whole projected Burgers circuit around the leading part was analysed $\mathbf{b}_{12,p} = \mathbf{F}_{12}\mathbf{S}_{12} = a/12[2\bar{1}1]$, which Burgers vector is regarded as $\mathbf{b}_{12} = a/6[112]$ (30° **Dγ**). Based on analysis, the reaction is one pure screw dislocation dissolution and one 60 mixed dislocation disassociation with two **CISFs**:

$$\underset{a/2\,[101]}{\mathbf{DA, 60°}} \rightarrow \underset{a/6[112]}{\mathbf{D\gamma, 30°}} + \text{CISF} + \underset{a/6[2\bar{1}1]}{\mathbf{\gamma A, 90°}} \tag{A.3.a}$$

$$\underset{a/2\,[011]}{\mathbf{DB, 0°}} \rightarrow \underset{a/6[112]}{\mathbf{D\gamma, 30°}} + \text{CISF} + \underset{a/6[\bar{1}21]}{\mathbf{\gamma B, 30°}} \tag{A.3.b}$$

where **DB** is a perfect pure screw dislocation, **DA** is a perfect 60° mixed dislocation, **Dγ** and **γB** are 30° Shockley partial dislocations, **γA** is 90° Shockley edge partial dislocations. The Shockley partial dislocations **Dγ** and **γB** are pinned at the the γ′/γ interface. The angle of Shockley partials **γA** and **γB** is 120° and therefore the leading partials will attract from each other to become a **Dγ**. Therefore, the reaction Eqs. (A.3) can be written as:

$$\underset{a/2\,[101]}{\mathbf{DA, 60°}} + \underset{a/2\,[011]}{\mathbf{DB, 0°}} \rightarrow \underset{a/3[112]}{\mathbf{2D\gamma, 30°}} + \text{SESF} + \underset{a/6[112]}{\mathbf{D\gamma, 30°}} \tag{A.4}$$

The reaction in Eq. (A.4) was ever reported in literature (*Vorontsov et.al. 2012*) before. The mode **SESF-3** has been reported and studied in a number of publicatons (*Barba et.al. 2017, Kovarik et.al. 2009, Smith 2016, Smith et.al. 2015*). The reaction is one pure screw dislocation dissolution and one 60 mixed dislocation disassociation with two **CISFs**:

$$\underset{a/2\,[\bar{1}10]}{\mathbf{AB, 60°}} \rightarrow \underset{a/6[\bar{2}1\bar{1}]}{\mathbf{A\gamma, 90°}} + \text{CISF} + \underset{a/6[\bar{1}21]}{\mathbf{\gamma B, 30°}} \tag{A.5.a}$$

$$\underset{a/2\,[011]}{\mathbf{DB, 0°}} \rightarrow \underset{a/6[112]}{\mathbf{D\gamma, 30°}} + \text{CISF} + \underset{a/6[\bar{1}21]}{\mathbf{\gamma B, 30°}} \tag{A.5.b}$$

where **DB** is a perfect pure screw dislocation, **AB** is a perfect 60° mixed full dislocation, **Aγ** is 90° edge Shockley partial dislocation, **Dγ** and **γB** are 30° mixed type Shockley partials. Similarlly, the angle of Shockley partials **Aγ** and **Dγ** is 120° and therefore the leading partials will attract from each other to become **γB**. The two leading partials (**γB**) have same sign and therefore these two leadings will repel from reach other. Therefore, there is one **SISF** bounded between two leading partials (**γB**). The reaction Eqs. (A.5) can be written as:

$$\underset{a/2[\bar{1}10]}{\mathbf{AB, 60°}} + \underset{a/2[011]}{\mathbf{DB, 0°}} \rightarrow \underset{a/6[\bar{1}21]}{\mathbf{\gamma B, 30°}} + \mathbf{SESF} + \underset{a/6[\bar{1}21]}{\mathbf{\gamma B, 30°}} + \mathbf{SISF} + \underset{a/6[\bar{1}21]}{\mathbf{\gamma B, 30°}} \tag{A.6}$$

**Fig. A.2** summarises 3 different SESF shearing modes in superalloys, mentioned above. The reaction Eqs. **(1), (3)** and **(5)** correponds to mode **SESF-1**, **SESF-2** and **SESF-3**.

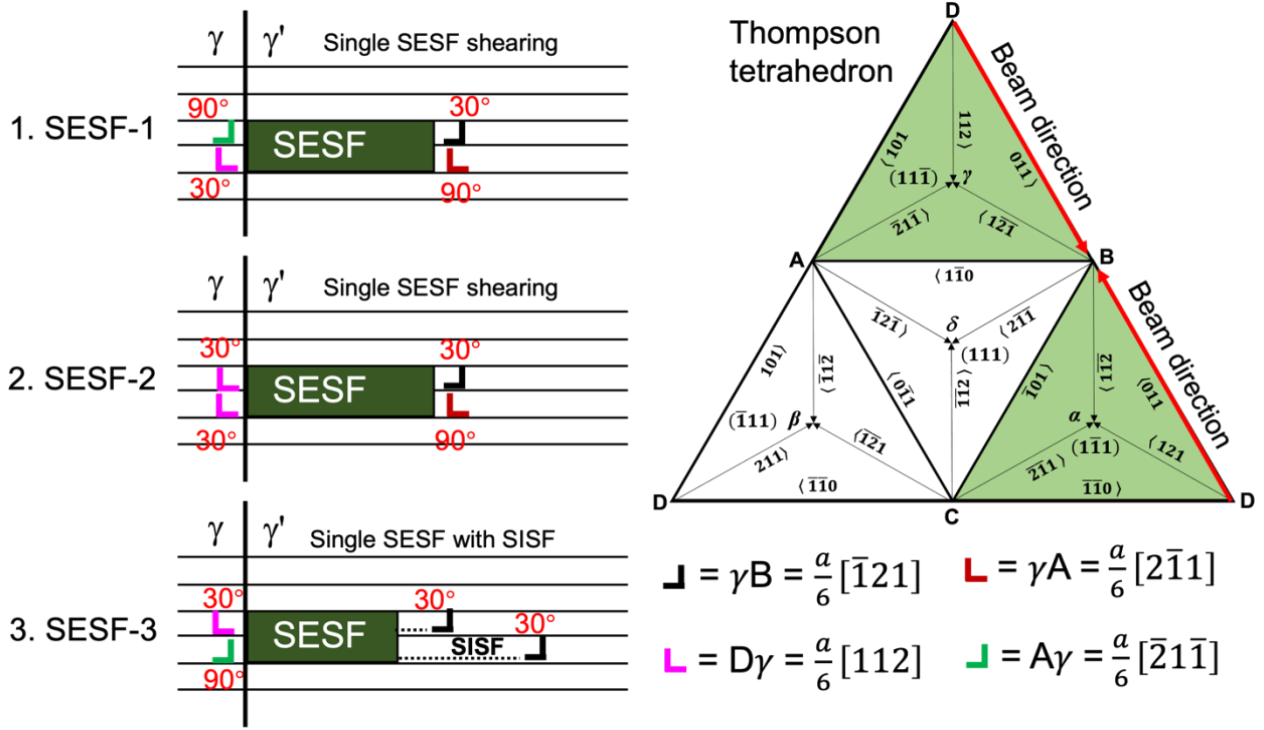

**Fig. A.2.** Summary of superlattice extrinsic stacking fault (SESF) shearing modes, i.e. SESF-1, SESF-2 and SESF-3 in superalloys.

**Appendix 2: Activation energy for 90° and 30° Shockley partial dislocation *cross-split* and cross-slip**

**Case 1: Activation energy for the "L-like" cross-slip with SISF**

For simplification, a two-dimensional model is adopted in this study, where the dislocation line is oriented perpendicular to the plane of the page. Based on this model, we can further calculate the energy of the unit dislocation. In the following section, the subscripts '2', '4' and '5' represent the dislocations **αB, αγ** and **γB,** respectively, as indicated in **Fig. A.3**. $r_i$ and $E_i$ denote the core radius of the dislocation *i* and its self-energy, separately. $r_{ij}$ ($r_{ij}$ is the corresponding magnitude) and $E_{ij}$ are the distance vector between the dislocations *i* and *j* and their interaction energy. It is apparent that $E_i$ and $E_{ij}$ can be calculated according to the following equations:

$$E_i = \frac{G(\boldsymbol{b}_i \cdot \boldsymbol{\mu})(\boldsymbol{b}_i \cdot \boldsymbol{\mu})}{4\pi} \ln \frac{r_a}{r_i} + \frac{G[(\boldsymbol{b}_i \times \boldsymbol{\mu})(\boldsymbol{b}_i \times \boldsymbol{\mu})]}{4\pi(1-\nu)} \ln \frac{r_a}{r_i} \tag{A.7}$$

$$E_{ij} = -\frac{G(\boldsymbol{b}_i \cdot \boldsymbol{\mu})(\boldsymbol{b}_i \cdot \boldsymbol{\mu})}{2\pi} \ln \frac{r_{ij}}{r_a} - \frac{G[(\boldsymbol{b}_i \times \boldsymbol{\mu})(\boldsymbol{b}_i \times \boldsymbol{\mu})]}{2\pi(1-\nu)} \ln \frac{r_{ij}}{r_a} - \frac{G[(\boldsymbol{b}_i \times \boldsymbol{\mu}) \cdot \boldsymbol{r}_{ij}][(\boldsymbol{b}_j \times \boldsymbol{\mu}) \cdot \boldsymbol{r}_{ij}]}{2\pi(1-\nu)r_{ij}^2} \tag{A.8}$$

where $G$ and $v$ denote the shear modulus and Poisson's ratio, respectively. $\boldsymbol{b}_i$ and $\boldsymbol{\mu}$ represent the Burgers vector and line direction of the dislocation $i$ and $j$, while $r_a$ is the outer cut-off radius, which can be canceled out in the following derivation. The line direction of all dislocations in this work is $\boldsymbol{\mu} = [011]$. The total energy $E_p$ per unit length for the dissociated configuration in **Fig. A.4** can be expressed as:

$$E_p = E_1 + E_2 + E_{12} + \gamma_{SISF} r_{12} \tag{A.9}$$

Here, we only consider the work done by the external stress on the cross-slip plane. After dislocation cross-slip, the total energy $E_p$ per unit length can be expressed:

$$E_c = E_1 + E_4 + E_3 + E_{14} + E_{13} + E_{34} + \gamma_{SISF} r_{34} + \gamma_{SISF} r_{14} - \tau b r_{34}/2 \tag{A.10}$$

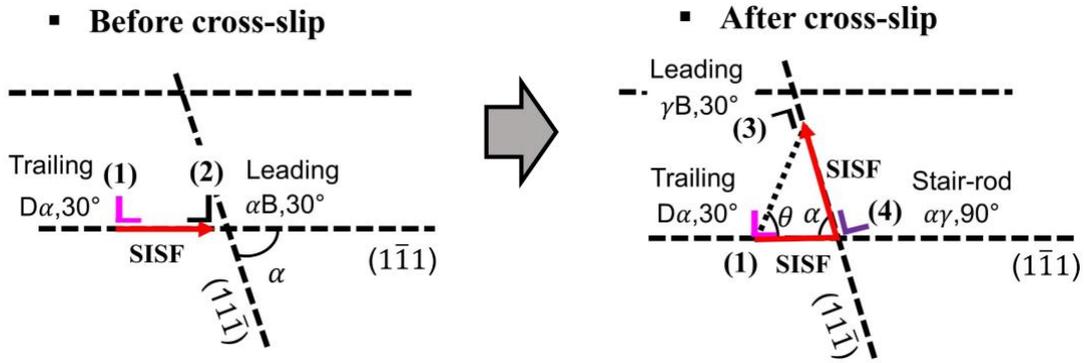

**Fig. A.3.** Case 1 - The Schematic model of the L-like cross slip for Schockley partial dislocation.

Table A1. The parameters related to our model.

| Symbol | Description |
|---|---|
| $G$ | Shear modulus |
| $\boldsymbol{\mu}$ | The dislocation direction |
| $v$ | Poisson's ratio |
| $\boldsymbol{b}_i$ | Burger vector of $i^{th}$ dislocation |
| $\boldsymbol{r}_{ij}$ | The distance vector between the dislocations $i$ and $j$ |
| $\tau$ | The applied stress |
| $r_i$ | Core radius of the dislocation $i$ |
| $\gamma_{SISF}, \gamma_{SIEF}$ | Stacking fault energy |
| $r_a$ | The outer cut-off radius |
| $E_i$ | Self energy of the $i^{th}$ dislocaiton |
| $E_{ij}$ | The interaction energy between the dislocations $i$ and $j$ |

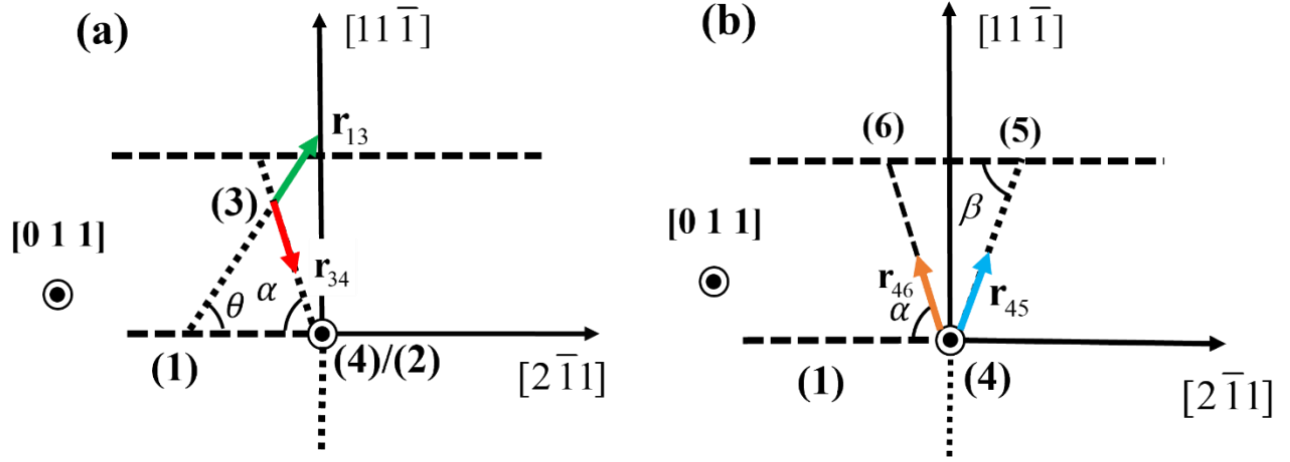

**Fig. A.4.** Angles associated with each vector during (a) the first cross-slip event, and (b) the first to the second cross slip event.

**Fig. A.4** illustrates the geometric relationships among the dislocation segments, where the angular parameters are used to define the relative position vectors. Specifically, **Fig. A.4(a)** corresponds to the dislocation geometry during the first cross-slip event in the L-like (or the initial stage of the Z-like) cross-slip mechanism. In contrast, **Fig. A.4(b)** depicts the configuration following the second cross-slip event in the Z-like cross-slip process. As shown in **Fig. A.4(a)**, we have:

$$\boldsymbol{r}_{13} = r_{13}\left[\frac{2\cos\theta}{\sqrt{6}} + \frac{\sin\theta}{\sqrt{3}}, -\frac{\cos\theta}{\sqrt{6}} + \frac{\sin\theta}{\sqrt{3}}, \frac{\cos\theta}{\sqrt{6}} - \frac{\sin\theta}{\sqrt{3}}\right] \quad (A.11)$$

$$\boldsymbol{r}_{34} = r_{34}\left[\frac{2\cos\alpha}{\sqrt{6}} - \frac{\sin\alpha}{\sqrt{3}}, -\frac{\cos\alpha}{\sqrt{6}} - \frac{\sin\alpha}{\sqrt{3}}, \frac{\cos\alpha}{\sqrt{6}} + \frac{\sin\alpha}{\sqrt{3}}\right] \quad (A.12)$$

$$\boldsymbol{r}_{14} = r_{14}\left[\frac{2}{\sqrt{6}}, -\frac{1}{\sqrt{6}}, \frac{1}{\sqrt{6}}\right]; \boldsymbol{r}_{12} = r_{12}\left[\frac{2}{\sqrt{6}}, -\frac{1}{\sqrt{6}}, \frac{1}{\sqrt{6}}\right] \quad (A.13)$$

And for the **Fig. A.4(b)**, we have:

$$\boldsymbol{r}_{45} = r_{45}\left[\frac{2\cos\beta}{\sqrt{6}} + \frac{\sin\beta}{\sqrt{3}}, -\frac{\cos\beta}{\sqrt{6}} + \frac{\sin\beta}{\sqrt{3}}, \frac{\cos\beta}{\sqrt{6}} - \frac{\sin\beta}{\sqrt{3}}\right] \quad (A.14)$$

$$\boldsymbol{r}_{46} = r_{46}\left[-\frac{2\cos\alpha}{\sqrt{6}} + \frac{\sin\alpha}{\sqrt{3}}, \frac{\cos\alpha}{\sqrt{6}} + \frac{\sin\alpha}{\sqrt{3}}, -\frac{\cos\alpha}{\sqrt{6}} - \frac{\sin\alpha}{\sqrt{3}}\right] \quad (A.15)$$

$$\boldsymbol{r}_{56} = r_{56}\left[-\frac{2}{\sqrt{6}}, \frac{1}{\sqrt{6}}, -\frac{1}{\sqrt{6}}\right] \quad (A.16)$$

From the geometric point of view in **Fig. A.3**, it is easy to obtain that $r_{12} = r_{14} = r_0$, and $E_1 = E_2 = E_3$. The dislocations (1), (2) and (3) are Shockley partial dislocations. Therefore, their line energy should be same. Based on energy variation in Eqs. (A.9) and (A.10), the activation energy barrier for cross-slip should be:

$$\Delta E = E_c - E_p = E_4 + E_{14} + E_{34} + E_{13} - E_{12} + \gamma_{SF} r_{34} - \tau b r_{34}/2 =$$

$$\frac{K_e b^2}{3}\left[\frac{1}{6}\cdot\ln\frac{r_{34}}{r_4}-\frac{1}{12}\ln\frac{r_0}{r_{13}}\right]+\frac{K_e b^2}{108}[\sqrt{2}\sin\alpha+3\sqrt{2}\sin2\theta+5\cos2\alpha-\frac{3}{2}\cos2\theta+\frac{9}{2}]$$

$$+\frac{K_s b^2}{4}\ln\frac{r_0}{r_{13}}-\left(\frac{\tau b}{2}-\gamma_{SISF}\right)r_{34} \qquad (A.17)$$

where $K_e=\frac{G}{2\pi(1-v)}$ and $K_s=\frac{G}{2\pi}$, $r_{13}=\sqrt{(r_0\sin\alpha)^2+(r_{34}-r_0\cos\alpha)^2}$, $\theta=arctan\left(\frac{r_{34}\sin\alpha}{r_0-r_{34}\cos\alpha}\right)$.

**Case 2: Activation energy for 90° Shockley and 90° Frank edge partial dislocation *cross-split* with SISF (Mode 1 for Z-like SF interaction explanation)**

In **Section 4.1**, **Mode 1** of the Z-like SF interaction involves two leading Shockley partials: a 90º Shockley and a 30º Shockley partial dislocation. As shown in **Fig. 10**, each of these leading dislocations precedes a SISF. The schematic model for the *cross-split* of edge partial dislocations is illustrated in **Fig. A.5**.

Eq. (A.18) describes the activation energy required for the **first** *cross-split* of the 90º Shockley edge partial. Eq. (A.19) accounts for the influence of solute segregation on this activation energy. Similarly, Eq. (A.20) defines the activation energy necessary for the **second** *cross-split* of the 90º Frank partial, while Eq. (A.21) incorporates the effect of element segregation on this second process.

**Fig. A.6** shows the relationship between activation energy and the required shear stress for the cross-splitting of the 90º leading partial in the SISF mode of the Z-like SF interaction. At the first splitting point, due to the inequality $|\mathbf{b}_{\gamma A}|^2<|\mathbf{b}_{\gamma\alpha}|^2+|\mathbf{b}_{\alpha A}|^2$, the activation energy is high, necessitating a correspondingly high applied shear stress. In contrast, at the second splitting point, where $|\mathbf{b}_{\alpha A}|^2>|\mathbf{b}_{\alpha\gamma}|^2+|\mathbf{b}_{\gamma A}|^2$, the required stress is reduced.

**Fig. A.7** presents the relationship between activation energy and shear stress for the cross-slip of the 30º leading partial dislocation in the same configuration. At the first cross-slip point, since $|\mathbf{b}_{\gamma B}|^2<|\mathbf{b}_{\gamma\alpha}|^2+|\mathbf{b}_{\alpha B}|^2$, a high activation energy and high shear stress are required. Likewise, at the second cross-slip point, the inequality $|\mathbf{b}_{\alpha B}|^2<|\mathbf{b}_{\alpha\gamma}|^2+|\mathbf{b}_{\gamma B}|^2$ also leads to a high activation energy and shear stress.

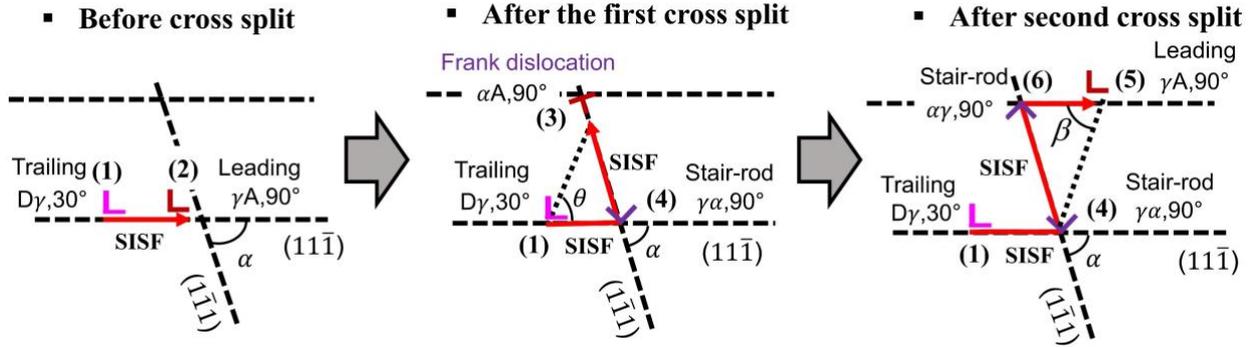

**Fig. A.5.** The Schematic model of the *cross-split* for edge partial dislocation in the Z-like SFs interaction. In the first cross point, the 90º Shockley edge partial was split into one stair-rod edge parial and one 90º Frank edge partial by "*cross-split* of edge dislocations", where the reaction energy increases $|b_{\gamma A}|^2 < |b_{\gamma \alpha}|^2 + |b_{\alpha A}|^2$. In the second cross point, the 90º Frank edge partial was split into one stair-rod edge parial and one 90º Shockley edge partial by "*cross-split* of edge dislocations", where the reaction energy decrease $|b_{\alpha A}|^2 > |b_{\alpha \gamma}|^2 + |b_{\gamma A}|^2$.

Similarly, we have $r_{14} = r_{12} = r_0$ for the *cross-split* model, shown in **Fig. A.5**. The $r_0$ in **Fig. 5(b)** is 108 nm. The activation energy required for the first 90º Shockley edge partial cross-split is:

$$\Delta E = -b^2 k_e \left[ -\frac{2}{9} \ln \frac{r_0}{r_{13}} + \frac{2}{9} \ln \frac{r_a}{r_{34}} + \frac{1}{6} \ln \frac{r_a}{r_2} - \frac{1}{3} \ln \frac{r_a}{r_3} - \frac{1}{18} \ln \frac{r_a}{r_4} \right.$$
$$\left. + \frac{5\sqrt{2}}{54} \sin 2\alpha + \frac{\sqrt{2}}{36} \sin 2\theta + \frac{1}{27} \cos 2\alpha - \frac{1}{9} \cos 2\theta \right] - r_{34}(\frac{\tau b}{2} - \gamma_{SISF}) \quad (A.18)$$

Note that the geometric configuration shown in the figure depends on the angles $\alpha$ and $\theta$. And, now we have $\alpha = 70.53°$, $r_{13} = \sqrt{(r_0 \sin\alpha)^2 + (r_{34} - r_0 \cos\alpha)^2}$ and $\theta = arctan\left(\frac{r_{34}\sin\alpha}{r_0 - r_{34}\cos\alpha}\right)$.

Under elements segregation consideration, the Eq. (A.18) can be expressed as:

$$\Delta E_C = -b^2 k_e \left[ -\frac{2}{9} \ln \frac{r_0}{r_{13}} + \frac{2}{9} \ln \frac{r_a}{r_{34}} + \frac{1}{6} \eta_1 \ln \frac{r_a}{r_2} - \frac{1}{3} \eta_1 \ln \frac{r_a}{r_3} - \frac{1}{18} \eta_1 \ln \frac{r_a}{r_4} \right.$$
$$\left. + \frac{5\sqrt{2}}{54} \sin 2\alpha + \frac{\sqrt{2}}{36} \sin 2\theta + \frac{1}{27} \cos 2\alpha - \frac{1}{9} \cos 2\theta \right] - r_{34}(\frac{\tau b}{2} - \eta_2 \gamma_{SISF}) \quad (A.19)$$

where $\eta_1 = 0.67$ is the coefficient for the elements segregation effect on dislocation line energy. $\eta_2 = 0.15$ is the coefficient for the elements segregation effect on SISF energy. The $\Delta U_{stir-Z} = -0.63 \times 10^{-10}$ J/m in Z-like SF interaction is the line energy reduction of a stair-rod dislocation ($\boldsymbol{b} = \frac{a}{6}[01\bar{1}]$) reducing energy by about 33%, compared with $0.5 G b^2$, where $G$ is 53.9 GPa at 850 °C. In order to simplify our mathematic model, the line energy reduction of all dislocations is considered as 33%.

For the second *cross-split*, we have $r_{34} = r_{46} = r_s$. The $r_0$ in **Fig. 5(b)** is measured as 150 nm. The activation energy required for the second 90° Frank edge partial *cross-split* is:

$$\Delta E = b^2 k_e \left[ -\frac{1}{3} \ln \frac{r_a}{r_3} - \frac{1}{9} \ln \frac{r_{56}}{r_{45}} + \frac{1}{6} \ln \frac{r_a}{r_5} + \frac{1}{9} \ln \frac{r_a}{r_s} + \frac{1}{18} \ln \frac{r_a}{r_6} \right.$$
$$\left. + \frac{\sqrt{2}}{18} \sin 2\alpha + \frac{\sqrt{2}}{18} \sin 2\beta + \frac{1}{18} \cos 2\alpha - \frac{1}{18} \cos 2\beta \right] - r_{56} \left( \frac{\tau b}{2} - \gamma_{SISF} \right) \quad (A.20)$$

And it holds, $r_{45} = \sqrt{(r_s \sin\alpha)^2 + (r_{56} - r_s \cos\alpha)^2}$ and $\beta = \arctan \left( \frac{r_s \sin\alpha}{r_{56} - r_s \cos\alpha} \right)$.

Under elements segregation consideration, the Eq. (A.20) can be expressed as:

$$\Delta E_C = b^2 k_e \left[ -\frac{1}{3} \eta_1 \ln \frac{r_a}{r_3} - \frac{1}{9} \ln \frac{r_{56}}{r_{45}} + \frac{1}{6} \eta_1 \ln \frac{r_a}{r_5} + \frac{1}{9} \ln \frac{r_a}{r_s} + \frac{1}{18} \eta_1 \ln \frac{r_a}{r_6} \right.$$
$$\left. + \frac{\sqrt{2}}{18} \sin 2\alpha + \frac{\sqrt{2}}{18} \sin 2\beta + \frac{1}{18} \cos 2\alpha - \frac{1}{18} \cos 2\beta \right] - r_{56} \left( \frac{\tau b}{2} - \eta_2 \gamma_{SISF} \right) \quad (A.21)$$

where $\eta_1 = 0.67$ is the coefficient for the elements segregation effect on dislocation line energy and $\eta_2 = 0.15$ is the coefficient for the elements segregation effect on SISF energy.

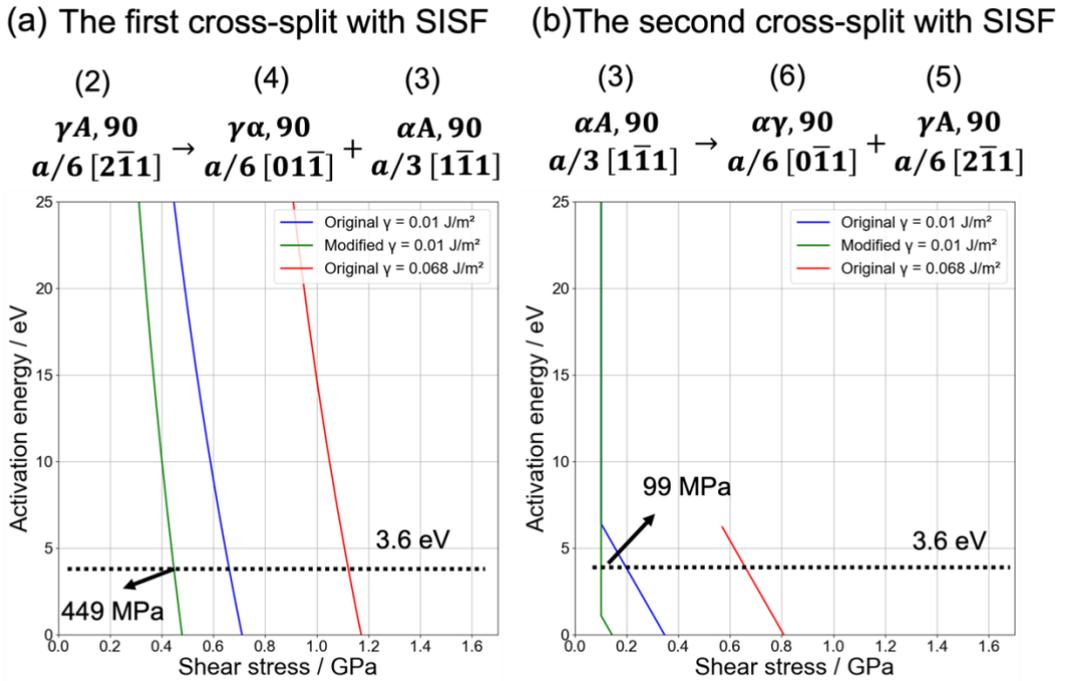

**Fig. A.6.** Relationship of **cross-split** activation energy and required shear stress in SISF mode: (a) First 90° Shockley edge partial *cross-split* and (b) The second 90° Frank edge partial *cross-split*. In the second *cross-split* event, the reaction energy decreases with $|b_{\alpha A}|^2 > |b_{\alpha \gamma}|^2 + |b_{\gamma A}|^2$. Therefore, the second *cross-split* event does not require high applied stress compared with the first *cross-split*.

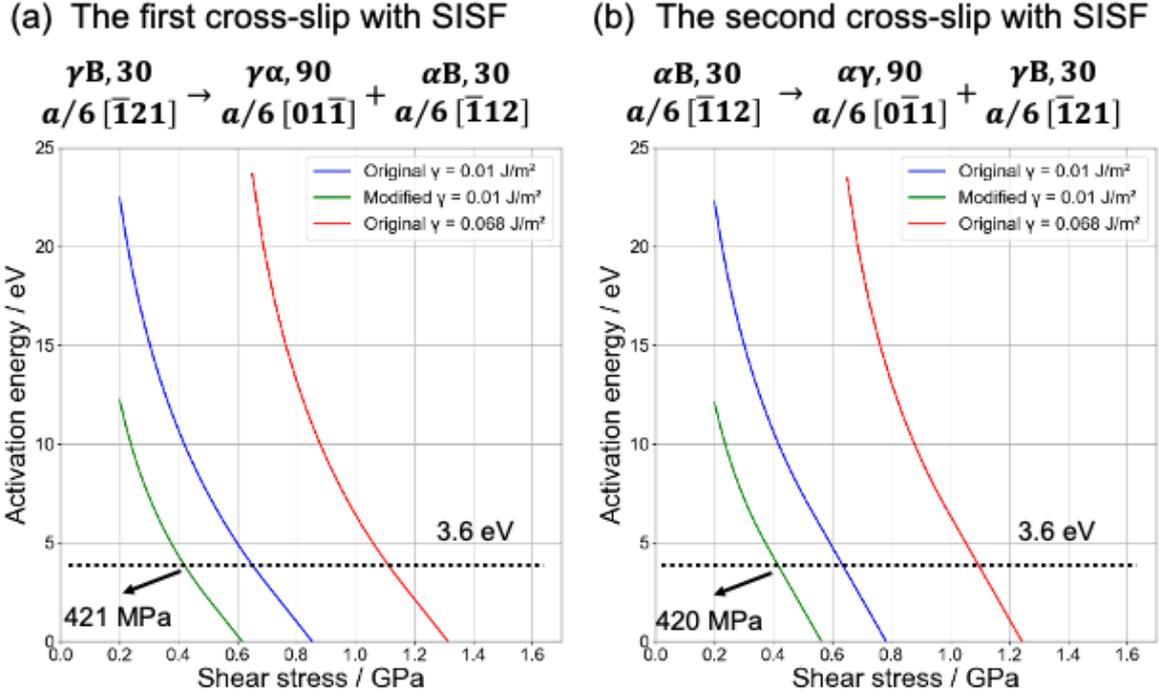

**Fig. A.7.** Relationship of **cross-slip** activation energy and required shear stress in SISF mode: (a) The first 30º Shockley partial cross-slip and (b) The second 30º Shockley partial cross-slip. There is no big difference in activation energy or necessary shear stress between the first cross-slip and the second cross-slip for 30º Shockley partials.

**Case 3: Activation energy for the 30° Shockley partial dislocation cross-slip with SESF (Mode 2 for Z-like SF interaction explanation)**

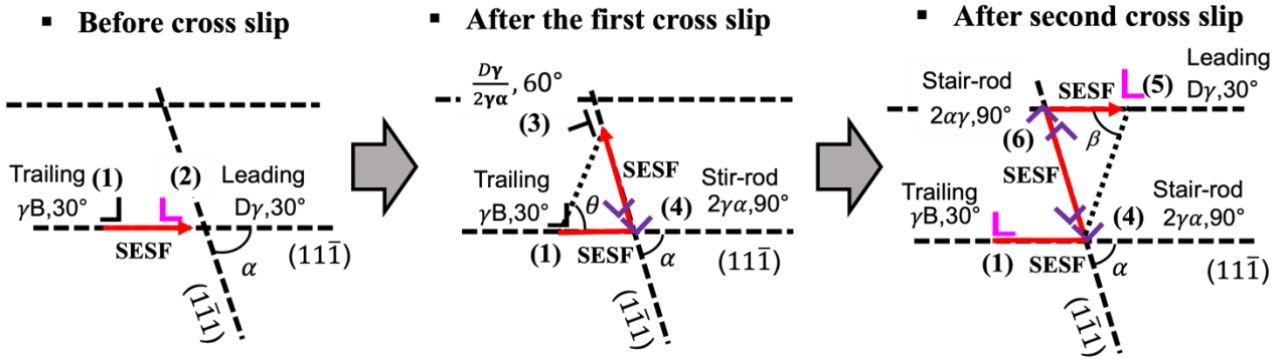

**Fig. A.8.** Case 3 - The Schematic model of the Z-like cross-slip for Shockley partial dislocation.

Similarly, we have $r_{14} = r_{12} = r_0 = 108 \, nm$ for the cross-slip model shown in **Fig. A.8**, the activation energy required for the first cross slip is:

$$\Delta E = b^2 k_e \left[ -\frac{7}{36} \ln\frac{r_0}{r_{13}} - \frac{1}{24} \ln\frac{r_a}{r_2} + \frac{3}{8} \ln\frac{r_a}{r_3} - \frac{5}{9} \ln\frac{r_a}{r_{34}} + \frac{2}{9} \ln\frac{r_a}{r_4} \right.$$

$$-\frac{11\sqrt{2}}{54}\sin 2\alpha + \frac{\sqrt{2}}{18}\sin 2\theta + \frac{1}{54}\cos 2\alpha - \frac{7}{72}\cos 2\theta + \frac{3}{8}\bigg] + \frac{b^2 k_s}{4}\ln\frac{r_0}{r_{13}} - r_{34}(\frac{\tau b}{2} - \gamma_{SESF}) \quad (A.22)$$

Note that the geometric configuration shown in **Fig. A.3** depends on the angles $\alpha$ and $\theta$. And, now we have $\alpha = 70.53°$. $r_{13} = \sqrt{(r_0\sin\alpha)^2 + (r_{34} - r_0\cos\alpha)^2}$ and $\theta = arctan\left(\frac{r_{34}\sin\alpha}{r_0 - r_{34}\cos\alpha}\right)$.

Under elements segregation consideration, the Eq. (A.22) can be expressed as:

$$\Delta E_C = b^2 k_e \bigg[-\frac{7}{36}\ln\frac{r_0}{r_{13}} - \frac{1}{24}\eta_1 \ln\frac{r_a}{r_2} + \frac{3}{8}\eta_1 \ln\frac{r_a}{r_3} - \frac{5}{9}\ln\frac{r_a}{r_{34}} + \frac{2}{9}\eta_1 \ln\frac{r_a}{r_4}$$

$$-\frac{11\sqrt{2}}{54}\sin 2\alpha + \frac{\sqrt{2}}{18}\sin 2\theta + \frac{1}{54}\cos 2\alpha - \frac{7}{72}\cos 2\theta + \frac{3}{8}\bigg] + \frac{b^2 k_s}{4}\ln\frac{r_0}{r_{13}} - r_{34}(\frac{\tau b}{2} - \eta_2\gamma_{SESF}) \quad (A.23)$$

where $\eta_1 = 0.67$ is the coefficient for the elements segregation effect on dislocation line energy. $\eta_2 = 0.15$ is the coefficient for the elements segregation effect on SESF energy. The SESF energy reduction caused by elements segregation in Z-like SF interaction is $\Delta\gamma_{SESF} = -0.071$ J/m$^2$, which is a reduction by about 80% compared with $\gamma_{SESF} = 0.089$ J/m$^2$ of L1$_2$ structured Ni$_3$Al (*Eurich et.al. 2015*).

For the second cross-slip, we have $r_{34} = r_{46} = r_s = 150$ nm. The activation energy required for the second cross slip is:

$$\Delta E = b^2 k_e \bigg[-\frac{3}{8}\ln\frac{r_a}{r_3} - \frac{1}{9}\ln\frac{r_{56}}{r_{45}} + \frac{1}{24}\ln\frac{r_a}{r_5} + \frac{1}{9}\ln\frac{r_a}{r_s} + \frac{2}{9}\ln\frac{r_a}{r_6}$$

$$+\frac{\sqrt{2}}{18}\sin 2\alpha + \frac{\sqrt{2}}{18}\sin 2\beta + \frac{1}{18}\cos 2\alpha - \frac{1}{18}\cos 2\beta\bigg] - r_{56}(\frac{\tau b}{2} - \gamma_{SESF}) \quad (A.24)$$

And it holds $r_{45} = \sqrt{(r_s\sin\alpha)^2 + (r_{56} - r_s\cos\alpha)^2}$. According to the geometric configuration shown in **Fig. A.4**, there is a relationship between the angles $\alpha$ and $\beta$ relationship that can be expressed by: $\beta = arctan\left(\frac{r_s\sin\alpha}{r_{56} - r_s\cos\alpha}\right)$.

Under elements segregation consideration, the Eq. (A.24) can be expressed as:

$$\Delta E_C = b^2 k_e \bigg[-\frac{3}{8}\eta_1 \ln\frac{r_a}{r_3} - \frac{1}{9}\ln\frac{r_{56}}{r_{45}} + \frac{1}{24}\eta_1 \ln\frac{r_a}{r_5} + \frac{1}{9}\ln\frac{r_a}{r_s} + \frac{2}{9}\eta_1 \ln\frac{r_a}{r_6}$$

$$+\frac{\sqrt{2}}{18}\sin 2\alpha + \frac{\sqrt{2}}{18}\sin 2\beta + \frac{1}{18}\cos 2\alpha - \frac{1}{18}\cos 2\beta\bigg] - r_{56}(\frac{\tau b}{2} - \eta_2\gamma_{SESF}) \quad (A.25)$$

where $\eta_1 = 0.67$ is the coefficient for the elements segregation effect on dislocation line energy. $\eta_2 = 0.2$ is the coefficient for the elements segregation effect on SESF energy.

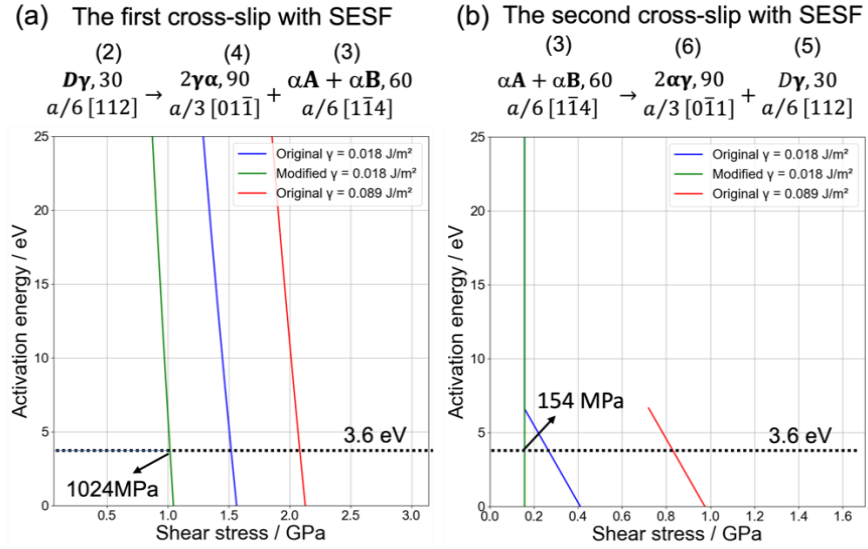

**Fig. A.9.** Relationship between cross-slip activation energy and required shear stress in SESF mode: (a) The first cross-slip and (b) The second cross-slip. In the first cross-slip event, the reaction energy increases significantly due to $|\mathbf{b}_{D\gamma}|^2 \ll |\mathbf{b}_{2\gamma\alpha}|^2 + |\mathbf{b}_{a/6[1\bar{1}4]}|^2$. Therefore, the first cross-slip event requires a quite high critical stress for this reaction, even under elements segregation effect. In the second cross-slip, the reaction energy decreases due to $|\mathbf{b}_{\alpha A}|^2 > |\mathbf{b}_{\alpha\gamma}|^2 + |\mathbf{b}_{\gamma A}|^2$. Therefore, the second cross-slip event does not require high applied stress compared with the first one.

**Declaration of generative AI and AI-assisted technologies in the writing process**

During the preparation of this work the author(s) used ChatGPT in order to improve writing. After using this tool/service, the author(s) reviewed and edited the content as needed and take(s) full responsibility for the content of the publication.

## 6. References

R.C. Reed, The superalloys: fundamentals and applications. Cambridge university press, (2008).

C.T. Sims, N.S. Stoloff, W.C. Hagel, Superalloys II: High-Temperature Materials for Aerospace and Industrial Power, J. Wiley & Sons, 1987, pp. 2-5.

Chatterjee, S., Li, Y., Po, G., 2021. A discrete dislocation dynamics study of precipitate bypass mechanisms in nickel-based superalloys. Int. J. Plast. 145, 103062.

Przybyla, C. P., McDowell, D. L., 2010. Microstructure-sensitive extreme value probabilities for high cycle fatigue of Ni-base superalloy IN100. Int. J. Plast. 26(3), 372-394.

Li, J., Chen, H., Fang, Q., Jiang, C., Liu, Y., Liaw, P. K., 2020. Unraveling the dislocation-precipitate interactions in high-entropy alloys. Int. J. Plast. 133, 102819.

Wu, R., Zaiser, M., Sandfeld, S., 2017. A continuum approach to combined γ/γ′ evolution and dislocation plasticity in Nickel-based superalloys. Int. J. Plast. 95, 142-162.

Nganbe, M., Heilmaier, M., 2009. High temperature strength and failure of the Ni-base superalloy PM 3030. Int. J. Plast. 25(5), 822-837.

Hou, J., Gan, J., Wang, T., Han, J., Ren, Z., Wang, Z., Yang, T., 2024. Dynamic strain ageing of L12-strengthened Ni-Co base high-entropy alloy and unraveling its deformation mechanisms in strain ageing process. Int. J. Plast.183, 104151.

Feng, L., Lv, D., Rhein, R. K., Goiri, J. G., Titus, M. S., Van der Ven, A., Wang, Y., 2018. Shearing of γ'particles in Co-base and Co-Ni-base superalloys. Acta Mater. 161, 99-109.

Chang, H. J., Fivel, M. C., Strudel, J. L., 2018. Micromechanics of primary creep in Ni base superalloys. Int. J. Plast. 108, 21-39.

Wang, S., Wu, W., Zhao, Y., Sun, Y., Song, C., Zhang, Y., Chen, H., 2025. Modulating $L1_2$ precipitation behavior and mechanical properties in an Fe-rich medium-entropy alloy fabricated via laser powder bed fusion. Int. J. Plast. *188*, 104290.

Zhang, Z., Ma, Y., Yang, M., Jiang, P., Feng, H., Zhu, Y., Yuan, F., 2024. Improving ductility by coherent nanoprecipitates in medium entropy alloy. Int. J. Plast. 172, 103821.

Borovikov, V. V., Mendelev, M. I., Smith, T. M., Lawson, J. W., 2023. Molecular dynamics simulation of twin nucleation and growth in Ni-based superalloys. Int. J. Plast. 166, 103645.

Borovikov, V. V., Mendelev, M. I., Smith, T. M., & Lawson, J. W., 2025. Stability of high energy superlattice faults in Ni-based superalloys from atomistic simulations. Int. J. Plast. 184, 104199.

Lenz, M., Eggeler, Y. M., Müller, J., Zenk, C. H., Volz, N., Wollgramm, P., Spiecker, E., 2019. Tension/Compression asymmetry of a creep deformed single crystal Co-base superalloy. Acta Mater.166, 597-610.

Titus, M. S., Eggeler, Y. M., Suzuki, A., & Pollock, T. M. (2015). Creep-induced planar defects in L12-containing Co-and CoNi-base single-crystal superalloys. Acta Mater. 82, 530-539.

Barba, D., Alabort, E., Pedrazzini, S., Collins, D. M., Wilkinson, A. J., Bagot, P. A. J., Reed, R. C., 2017. On the microtwinning mechanism in a single crystal superalloy. Acta Mater.135, 314-329.

L. Kovarik, R.R. Unocic, J. Li, P. Sarosi, C. Shen, Y. Wang, M.J. Mills, Microtwinning and other shearing mechanisms at intermediate temperatures in Ni-based superalloys. Prog. Mater. Sci. 54(6) (2009) 839-873.

T.M. Smith Jr, Orientation and alloying effects on creep strength in Ni-based superalloys (Doctoral dissertation, The Ohio State University), (2016).

N. Tsuno, S. Shimabayashi, K. Kakehi, C.M.F. Rae, R.C. Reed. Tension/Compression asymmetry in yield and creep strengths of Ni-based superalloys. In Superalloys 2008, Proceedings of the International Symposium on Superalloys, pages 433-442, 2008.

Lu, S., Antonov, S., Li, L., Liu, C., Zhang, X., Zheng, Y., Feng, Q., 2020. Atomic structure and elemental segregation behavior of creep defects in a Co-Al-W-based single crystal superalloys under high temperature and low stress. Acta Mater.190, 16-28.

Qi, D., Fu, B., Du, K., Yao, T., Cui, C., Zhang, J., Ye, H., 2016. Temperature effects on the transition from Lomer-Cottrell locks to deformation twinning in a Ni-Co-based superalloy. Scripta Mater.125, 24-28.

Jin, C., Xiang, Y., Lu, G., 2011. Dislocation cross-slip mechanisms in aluminum. Philos. Mag. 91(32), 4109-4125.

Le Hazif, R., Poirer, J. P., 1975. Cross-slip on {110} planes in aluminum single crystals compressed along <100> axis. Acta Metall. 23(7), 865-871.

Watanabe, C., Yamazaki, S., Koga, N., 2021. Effects of cross-slip activity on low-cycle fatigue behavior and dislocation structure in pure aluminum single crystals with single-slip orientation. Mater. Sci. Eng. A 815, 141221.

Bonneville, J., Escaig, B., Martin, J. L., 1988. A study of cross-slip activation parameters in pure copper. Acta Metall. 36(8), 1989-2002.

Rasmussen, T., Jacobsen, K. W., Leffers, T., Pedersen, O. B., 1997. Simulations of the atomic structure, energetics, and cross slip of screw dislocations in copper. Physical Review B, 56(6), 2977.

Rao, S. I., Dimiduk, D. M., El-Awady, J. A., Parthasarathy, T. A., Uchic, M. D., Woodward, C., 2010. Activated states for cross-slip at screw dislocation intersections in face-centered cubic nickel and copper via atomistic simulation. Acta Materi. 58(17), 5547-5557.

Kang, K., Yin, J., Cai, W., 2014. Stress dependence of cross slip energy barrier for face-centered cubic nickel. J. Mech. Phys. Solids. 62, 181-193.

Kuykendall, W. P., Wang, Y., Cai, W., 2020. Stress effects on the energy barrier and mechanisms of cross-slip in FCC nickel. J. Mech. Phys. Solids. 144, 104105.

Wen, M., Fukuyama, S., Yokogawa, K., 2004. Hydrogen-affected cross-slip process in fcc nickel. Physical Review B, 69(17), 174108.

Lin, T. L., Wen, M., 1990. The deformation mechanism of a γ′ precipitation-hardened nickel-base superalloy. Mater. Sci. Eng. A 128(1), 23-31.

Milligan, W. W., Antolovich, S. D., 1989. On the mechanism of cross slip in $Ni_3Al$. Metall. Mater. Trans. A, 20, 2811-2818.

Rai, R. K., Sahu, J. K., Paulose, N., Fernando, D. C., 2021. Tensile deformation micro-mechanisms of a polycrystalline nickel base superalloy: from jerky flow to softening. Mater. Sci. Eng. A 807, 140905.

Zhang, J. X., Wang, J. C., Harada, H., Koizumi, Y., 2005. The effect of lattice misfit on the dislocation motion in superalloys during high-temperature low-stress creep. Acta Mater. 53(17), 4623-4633.

León-Cázares, F. D., Schlütter, R., Monni, F., Hardy, M. C., Rae, C. M., 2022. Nucleation of superlattice intrinsic stacking faults via cross-slip in nickel-based superalloys. Acta Mater. 241, 118372.

Qi, D., Fu, B., Du, K., Yao, T., Cui, C., Zhang, J., Ye, H., 2016. Temperature effects on the transition from Lomer-Cottrell locks to deformation twinning in a Ni-Co-based superalloy. Scripta Mater. 125, 24-28.

Behjati, P., Asgari, S., 2011. Microstructural characterisation of deformation behaviour of nickel base superalloy IN625. Mater. Sci. Technol. 27(12), 1858-1862.

Zeng, S., Zhao, X., Xia, W., Fan, Y., Qiao, L., Cheng, Y., Zhang, Z., 2024. Effect of temperature on the tensile deformation mechanisms of a fourth nickel-based single crystal superalloys. J. Mater. Res. Technol. 30, 6254-6264.

Zhou, H. J., Chang, H., Feng, Q., 2017. Transient minimum creep of a γ′ strengthened Co-base single-crystal superalloy at 900° C. Scripta Mater. 135, 84-87.

Messé, O. M., Barnard, J. S., Pickering, E. J., Midgley, P. A., Rae, C. M. F., 2014. On the precipitation of delta phase in ALLVAC® 718Plus. Philos. Mag. 94(10), 1132-1152.

Karpstein, N., Lenz, M., Bezold, A., Wu, M., Neumeier, S., Spiecker, E., 2023. Reliable identification of the complex or superlattice nature of intrinsic and extrinsic stacking faults in the L1$_2$ phase by high-resolution imaging. Acta Mater. 260, 119284.

Borovikov, V. V., Mendelev, M. I., Zarkevich, N. A., Smith, T. M., Lawson, J. W., 2024. Effect of Nb solutes on the Kolbe mechanism for microtwinning in Ni-based superalloys. Int. J. Plast. 178, 104004.

Vorontsov, V. A., Kovarik, L., Mills, M. J., Rae, C. M. F., 2012). High-resolution electron microscopy of dislocation ribbons in a CMSX-4 superalloy single crystal. Acta Mater. 60(12), 4866-4878.

Smith, T. M., Esser, B. D., Antolin, N., Viswanathan, G. B., Hanlon, T., Wessman, A., Mills, M. J., 2015. Segregation and η phase formation along stacking faults during creep at intermediate temperatures in a Ni-based superalloy. Acta Mater. 100, 19-31.

Smith, T.M., Esser, B.D., Good, B., 2018. Segregation and Phase Transformations Along Superlattice Intrinsic Stacking Faults in Ni-Based Superalloys. Metall. Mater. Trans. A 49, 4186-4198.

Eurich, N. C., Bristowe, P. D., 2015. Segregation of alloying elements to intrinsic and extrinsic stacking faults in γ′-Ni$_3$Al via first principles calculations. Scripta Mater. 102, 87-90.

Kositski, R., Kovalenko, O., Lee, S. W., Greer, J. R., Rabkin, E., Mordehai, D., 2016. Cross-split of dislocations: an athermal and rapid plasticity mechanism. Sci. Rep. 6(1), 25966.

Langmuir, I., 1918. The adsorption of gases on plane surfaces of glass, mica and platinum. J. Am. Chem. Soc. 40(9), 1361-1403,

McLean D., 1957. Grain Boundaries in Metals. Monographs on the physics and chemistry of materials. Clarendon Press.

Murdoch H. A., Schuh C. A., 2014. Estimation of grain boundary segregation enthalpy and its role in stable nanocrystalline alloy design. Journal of Materials Research, 28(16), 2154-2163.

Wagih M., Larsen P. M., Schuh C. A., 2020. Learning grain boundary segregation energy spectra in polycrystals. Nat. Commun. 11(1):6376, 2020.

Trelewicz J. R., Schuh C. A., 2009. Grain boundary segregation and thermodynamically stable binary nanocrystalline alloys. Physical Review B 79(9), 094112.

Kirchheim, R., 2007. Reducing grain boundary, dislocation line and vacancy formation energies by solute segregation: II. Experimental evidence and consequences. Acta Mater. 55(15), 5139-5148.

Medouni, I., Portavoce, A., Maugis, P., Eyméoud, P., Yescas, M., Hoummada, K. 2021. Role of dislocation elastic field on impurity segregation in Fe-based alloys. Sci. Rep. 11(1), 1780.

Bishop G. H., Chalmers B., 1968. A coincidence-ledge-dislocation description of grain boundaries. Scripta Metall. 2(2), 133-139.

Hirth J. P., Balluffi R. W., 1973. On grain boundary dislocations and ledges. Acta Metall. 21(7), 929-942.

Meiners T., Frolov T., Rudd R. E., Dehm G., Liebscher C. H., 2020. Observations of grain-boundary phase transformations in an elemental metal. Nature, 579(7799), 375-378.

Li, K. Q., Zhang, Z. J., Yan, J. X., Yang, J. B., Zhang, Z. F., 2018. Competition between two Fleischer modes of cross slip in silver. Computational Materials Science, 152, 93-98.

Galindo-Nava, E. I., Connor, L. D., Rae, C. M. F., 2015. On the prediction of the yield stress of unimodal and multimodal γ′ Nickel-base superalloys. Acta Mater. 98, 377-390.

Rao, Y., Smith, T. M., Mills, M. J., Ghazisaeidi, M., 2018. Segregation of alloying elements to planar faults in γ′-$Ni_3Al$. Acta Mater. 148, 173-184.

Li, Y., Chatterjee, S., Martinez, E., Ghoniem, N., Po, G., 2021. On the cross-slip of screw dislocations in zirconium. Acta Mater. 208, 116764.

Hussein, A. M., Rao, S. I., Uchic, M. D., Dimiduk, D. M., El-Awady, J. A., 2015. Microstructurally based cross-slip mechanisms and their effects on dislocation microstructure evolution in fcc crystals. Acta Mater. 85, 180-190.

Geipel, T., Xiao, S. Q., Pirouz, P., 1993. HRTEM study of 1/6⟨411⟩ partial dislocations in hot hardness indented Ge. Philos. Mag. Lett. 67(4), 245–251.

Xue, F., Zenk, C. H., Freund, L. P., Hoelzel, M., Neumeier, S., Göken, M., 2018. Double minimum creep in the rafting regime of a single-crystal Co-base superalloy. Scripta Mater.142, 129-132.